\documentclass[10pt,aps,prc,twocolumn,superscriptaddress,showpacs,floatfix]{revtex4-2}
\DeclareUnicodeCharacter{2062}{}

\usepackage{graphics}
\usepackage{graphicx}
\usepackage{dcolumn}
\usepackage{longtable}
\usepackage{amsmath}
\usepackage{multirow}
\usepackage{color}
\usepackage{sidecap}
\usepackage{wrapfig}
\usepackage[utf8]{inputenc}
\usepackage{newunicodechar}
\newunicodechar{γ}{\textgamma}

\usepackage{booktabs}
\usepackage{threeparttable}
\usepackage{standalone}
\usepackage{booktabs, dcolumn}
\usepackage{mathtools}
\usepackage{gensymb}
\usepackage[english]{babel}
\usepackage[utf8]{inputenc}
\usepackage{physics}
\usepackage{afterpage}
\usepackage{mathtools}
\usepackage{nccmath}
\usepackage{amssymb}
\usepackage{textcomp}

\date{\today}

\begin{document}

\title{Interplay of quadrupole and octupole degrees of freedom in the Gd isotopes}
\author{R. Budaca}
\email{rbudaca@theory.nipne.ro}
\affiliation{Horia Hulubei National Institute for R\&D in Physics and Nuclear Engineering, Str. Reactorului 30, RO-077125, POB-MG6 Bucharest-M\v{a}gurele, Romania}
\affiliation{Academy of Romanian Scientists, 54 Splaiul Independen\c{t}ei, RO-050094, Bucharest, Romania}

\author{S. Pascu}
\email{sorin.pascu@nipne.ro}
\affiliation{Horia Hulubei National Institute for R\&D in Physics and Nuclear Engineering, Str. Reactorului 30, RO-077125, POB-MG6 Bucharest-M\v{a}gurele, Romania}

\begin{abstract}
A systematic theoretical investigation of the quadrupole and octupole collective properties across the Gd isotopic chain is performed employing a quadrupole-octupole axially symmetric model. These nuclei have recently attracted significant attention following the revelation that the maximum octupole collectivity in this region is located at $^{150}$Gd. The model parameters are optimized by fitting to the low-lying positive and negative-parity energy levels, as well as to known $E0$, $E1$, $E2$, and $E3$ transition strengths. Our primary objective is a simultaneous and unified description of quadrupole and octupole collectivity across the even-even Gd nuclei in the $84\leqslant N \leqslant96$ range, a region that includes the transition from spherical to rotational nuclear shapes. The results show a smooth evolution of the quadrupole deformation, highlighted by a distinct jump at the well-known $N=90$ critical point. The enhancement of quadrupole deformation is also correlated with the loss of non-zero octupole deformation, which is reported only for the lightest $^{148,150}$Gd nuclei. This translates into a fair agreement with the measured $E3$ strength, predicting a maximum $B(E3)$ value for the $^{152}$Gd isotope.
\end{abstract}


\maketitle

\section{Introduction}
\label{Introduction}
The atomic nucleus is a complex, strongly interacting many-body quantum system whose structure is governed by the interplay of collective and single-particle degrees of freedom. While the theoretical studies have made significant progress in recent years, allowing for a better understanding of the intricate excitation modes that appear even at low energy, they still offer a relatively simplified version of reality. However, with the addition of new experimental data, new ideas can be tested and the models improved. 

One of the regions where a wealth of structural paradigms appears is the rare-earth region spanning the $50\leqslant Z\leqslant 82$, $82\leqslant N\leqslant 126$ part of the nuclear chart, offering a unique laboratory for investigating the emergence and coexistence of various collective phenomena in both even-even and odd-mass nuclei. In particular, Gd isotopes ($Z = 64$) exhibit a diverse range of structural phenomena, which pose a challenge to any theoretical model aimed at describing their subtle particularities. 

Although $Z = 64$ is not considered a complete shell closure, simple observables such as the energy of the first state $2^{+}_{1}$ or the $E(4^{+}_{1})/E(2^{+}_{1})$ ratio have revealed a so-called “bubble” pattern \cite{Cakirli2008}, which is indicative of a sub-shell closure. Additional proofs on the “magicity” of $Z = 64$ come from analyzing single-particle states around $^{146}$Gd \cite{Kleinheinz1979}, from a sharp discontinuity in the two-proton separation energies \cite{Blomqvist1983}, or from an abrupt change in charge radii observed in this region \cite{Ahmad1985}. Starting from these observations, it follows that $^{146}$Gd ($N = 82$) should exhibit properties of a doubly magic nucleus. Since traditionally $^{208}$Pb is considered the best example of a doubly magic nucleus, $^{146}$Gd was deemed as having a particular importance and a comparison between the two nuclei was performed \cite{Walet1989}. The calculations within the large-scale shell model show that while $^{146}$Gd clearly exhibits doubly magic characteristics, its stability against particle-hole excitations is not as pronounced as for the $^{208}$Pb, and propose a less pronounced shell closure at $Z = 64$ than at $Z = 82$ \cite{Walet1989}. 

In a mean-field description of the nucleus, octupole correlations arise when single-particle orbitals with $\Delta j = \Delta l = 3$ come close in energy. This situation is expected at the standard octupole magic numbers, near $N\simeq$ 34, 56, 88, and 134 \cite{Butler2016}, and is attributed to the odd-multipolarity neutron–proton (isoscalar) component of the nuclear interaction \cite{Chen2021}. Theoretical calculations with the proxy-SU(3) model have found that the octupole magic numbers are 32, 56, 90, 134, and 194 \cite{Martinou2024}. Among them, one can notice a robust prediction of enhanced octupole collectivity in the region of $N=88-90$, which in the case of Gd isotopes would correspond to $^{152}$Gd. The present study aims to confirm this behavior.

Around $N = 90$, the rare-earth region is also home to the so-called $X(5)$ symmetry introduced by Iachello in Ref. \cite{Iachello2001}. This simple parameter-free model serves as a critical point of the shape-phase transition from the vibrational regime to the rotational character. Besides $^{152}$Sm \cite{Casten2001} and $^{150}$Nd \cite{Krucken2002}, considered to be the best realizations of this symmetry, $^{154}$Gd \cite{Tonev2004} has also been proposed as a very good candidate based on comparison with energy levels and transition probabilities. However, since in $X(5)$ the potential used is an infinite square well, other theoretical studies have looked for the signature of a flat potential in their potential energy surfaces (PES). Interestingly, theoretical calculations performed with the relativistic mean-field approach in \cite{Sheng2005} and with the Hartree-Bogoliubov model in \cite{Fossion2006} have proposed that the best critical point symmetry is realized for $^{150}$Gd and $^{152}$Gd isotopes, and not for $^{154}$Gd. Many other theoretical investigations can be found in the literature on the subject of shape-phase transitions \cite{Shi2018, Naz2018, Quan2018, Guzman2007}. Before we conclude this part, we note three more studies. The first investigation was performed with the Monte Carlo shell model (MCSM) for Nd and Sm isotopes in this region. Remarkably, in the case of $^{150}$Nd and $^{150}$Sm, the authors found two different deformations in the ground state wave functions and concluded the existence of a “double-prolate-shape coexistence with mixing” \cite{Tsunoda2023}. The second one employed the constrained mean-field interacting boson model calculations for Nd, Sm, and Dy isotopes, and inferred the existence of the critical point symmetry from the behavior of the two-neutron transfer intensities in $(p,t)$ and $(t,p)$ reactions \cite{Nomura2019}. Finally, the last one employed the mean field and the Gogny D1S interaction, revealing that none of the calculated potential surfaces exhibit the structure of the critical point (a flat PES) \cite{Rodriguez2009}. Instead, the triaxial degree of freedom begins to play a significant role here.

Moving to even heavier Gd isotopes, large quadrupole moments have been confirmed from Coulomb excitation experiments for $^{156-160}$Gd \cite{Reich2012,Nica2017,Nica2021}, pointing to large deformation parameters around $\beta_2\simeq$ 0.35. We are therefore entering a region of strongly axially deformed nuclei. However, this conventional picture was recently challenged by the new MCSM calculations, which predict a large occurrence of triaxial shapes in this region \cite{Otsuka2025}. Gd nuclei are no exception, and triaxiality parameters of about 6$\degree$ are predicted for these nuclei. In fact, MCSM calculations are not the only ones predicting an increased triaxiality in this region. A new picture emerges following systematic calculations with the proxy-$SU(3)$ \cite{Bonatsos2024}, pseudo-$SU(3)$ \cite{Vargas2013}, and the triaxial projected shell model \cite{Rouoof2024}.

A recent measurement of the $E3$ strength has revealed that the octupole collectivity is reaching a maximum value for $^{150}$Gd ($N = 86$) \cite{Pascu2025}. With a $B(E3)$ value of 45(5) W.u. \cite{Pascu2025}, corroborated with preliminary indications that the $E3$ strength in the neutron-rich $^{144}$Ba (we note that this nucleus is a doubly magic nucleus with respect to the octupole deformation) is in fact less than half \cite{Jones2025} the published value of 48$^{+25}_{-34}$ W.u. \cite{Bucher2016}, this makes the Gd nuclei the isotopic chain with the strongest octupole collectivity in this region. The large $B(E3)$ values start already with the doubly magic $^{146}$Gd. In fact, a direct comparison with $^{208}$Pb reveals that both nuclei have as the first excited level a $L^{\pi} = 3^{-}$ state, which decays via a collective $E3$ transition with a $B(E3)$ value of 37(4) W.u. for $^{146}$Gd and with 34(1) W.u. for $^{208}$Pb. The increase of $E3$ strength continues then with $^{148}$Gd and culminates with $^{150}$Gd, while the $B(E3)$ values for heavier nuclei starting with $^{154}$Gd are about half (or less) than the ones in the $N = 82-86$ region. The increase of the $E3$ strength at the beginning of the shell was best reproduced in \cite{Pascu2025} by the quasi-particle random phase approximation (QRPA) with the Skyrme-type SkX interaction. The main configurations found responsible for this behavior were the two quasi-particle configurations $\pi1h_{11/2}-\pi2d_{5/2}$ and $\pi1h_{11/2}-\pi1g_{7/2}$. While the study of the Gd isotopic chain was approached with a group of state-of-the-art theoretical models in \cite{Pascu2025}, only a partial reproduction of the octupole collectivity could be obtained in the case of spherical nuclei. This demonstrates once again that the simultaneous reproduction of quadrupole and octupole degrees of freedom across an entire isotopic chain remains a challenge for current theoretical approaches.   

The presence of both quadrupole and octupole modes at low energies in Gd isotopes offers a strict testing ground for modern nuclear models. With a vibrational-like effect due to the increased octupolarity and the purely quadrupole shapes, separated by only a few hundred keV \cite{Guzman2023}, this structure provides a unique opportunity to study the reflection symmetry breaking and quadrupole collectivity within a single isotopic chain. In turn, having five stable even-even isotopes, Gd nuclei are experimentally accessible with very high precision through a multitude of methods, including Coulomb excitation and lifetime measurements, which enable the extraction of the relevant matrix elements. 

In this study, we have performed a systematic investigation of the even-even nuclei in the Gd isotopic chain in the $N = 84$ to 96 region using a quadrupole-octupole axially symmetric collective model. Our aim is to reproduce simultaneously the structure associated with quadrupole and octupole collectivity. In Section \ref{Theoretical framework}, we present the theoretical framework for the model Hamiltonian and the electromagnetic observables. Section \ref{Numerical_application_and_discussions} presents details of the numerical calculations, with a detailed comparison of the results with the experimental data. The final conclusions are drawn in Section \ref{Conclusions}.

\section{Theoretical framework}
\label{Theoretical framework}
\subsection{Model Hamiltonian}

For a combined reproduction of the rotational sequences and of the parity splitting energy, we consider the following Hamiltonian:
\begin{equation}
\hat{H}=\hat{H}_{QO}+C\hat{L}^{2},
\end{equation}
where $\hat{L}$ is the total angular momentum operator. The part responsible for the collective fluctuations of the quadrupole ($\beta_{2}$) and octupole ($\beta_{3}$) deformation is described by means of an axially symmetric quadrupole-octupole Bohr Hamiltonian \cite{Denisov1995}:
\begin{eqnarray}
\hat{H}_{QO}&=&-\sum_{\lambda=2,3}\frac{\hbar^{2}}{2B_{\lambda}}\frac{1}{\beta^{3}_{\lambda}}\frac{\partial}{\partial{\beta}_{\lambda}}\beta^{3}_{\lambda}
\frac{\partial}{\partial{\beta}_{\lambda}}\label{quoc}\\
&&+\frac{\hbar^{2}\hat{L}^{2}}{6(B_{2}\beta^{2}_{2}+2B_{3}\beta^{2}_{3})}+U(\beta_{2},\beta_{3}).\nonumber
\end{eqnarray}
The rotation-vibration dynamics of this operator is determined by the $\lambda$-pole mass parameters and the collective potential $U$. A new pair of variables
\begin{equation}
\tilde{\beta}=\sqrt{\frac{2(B_{2}\beta^{2}_{2}+B_{3}\beta^{2}_{3})}{B_{2}+B_{3}}},\,\,\,\tan{\phi}=\frac{\beta_{3}}{\beta_{2}}\sqrt{\frac{B_{3}}{B_{2}}},
\end{equation}
is more convenient for analytical purposes. The angular variable $\phi\in(-\pi/2,\pi/2)$ describes the relative contribution of the quadrupole and octupole deformation variables, while $\tilde{\beta}\ge0$ acts as an overall deformation measure. Note also that the $\pm\phi$ values correspond to the $\pm\beta_{3}$ deformation. A $\phi$-independent potential allows a factorized solution $\Phi^{\pm}_{LM0}(\tilde{\beta},\phi,\Omega)=\psi^{\pm}_{L}(\tilde{\beta})\chi^{\pm}_{L}(\phi)|LM0\pm\rangle$, where $M$ and $K=0$ are projections of the angular momentum on the $z$ axis of the laboratory and of the intrinsic reference frames. The parity of the total wave function is related to its transformation properties with respect to the reflection of the system in a plane normal to the symmetry axis \cite{Denisov1995,Bonatsos2005}.

The integration over the Euler angles $\Omega$ of the rotational function $|LM_{L}0\pm\rangle$, after the separation of the variables, provides a radial-like equation for the $\tilde{\beta}$ variable. It contains a separation constant $W^{\pm}_{L}$ which is defined as the eigenvalue of the angular equation
\begin{equation}
\left[-\frac{\partial^{2}}{\partial{\phi}^{2}}+u_{L}({\phi})\right]\chi^{\pm}_{L}(\phi)=W^{\pm}_{L}\chi^{\pm}_{L}(\phi).
\label{eqphi}
\end{equation}
The potential
\begin{equation}
u_{L}(\phi)=\frac{3}{\sin^{2}{2\phi}}+\frac{L(L+1)}{3(1+\sin^{2}{\phi})}
\label{uLphi}
\end{equation}
is intrinsic to the model's geometry and is changed as
\begin{equation}
\tilde{u}_{L}(\phi)=\left\{\begin{array}{l}v_{L}(\phi)=-a\phi^{2}+b,\,\,|\phi|<|\phi_{c}|,\\
u_{L}(\phi),\,\,|\phi|\ge|\phi_{c}|,\end{array}\right.
\label{poti}
\end{equation}
in order to allow interaction between $\pm\phi(\pm\beta_{3})$ configurations \cite{Budaca2022}. The inner parabola is parametrized by the continuity conditions at an adjustable connection value $|\phi_{c}|$. The new potential can thus accommodate single or double non-zero minima, consistent with the vibration or the stability of the octupole deformation \cite{Budaca2024}. The corresponding angular solutions are obtained through a diagonalization in a basis of particle-in-the-box wave functions.

The remaining $\tilde{\beta}$ equation is solved for a harmonic oscillator potential $\tilde{\beta}^{2}$, with an adjustable centrifugal contribution $w_{0}/\tilde{\beta}^{2}$. The total energy can then be expressed as
\begin{equation}
E_{Ln}^{\pm}=\frac{2\hbar^{2}}{B_{2}+B_{3}}\left[2n+\nu^{\pm}_{L}+1+2cL(L+1)\right],\label{Etot}
\end{equation}
where $c=C(B_{2}+B_{3})/(2\hbar^{2})$ and $\nu^{\pm}_{L}=\sqrt{W_{L}^{\pm}+w_{0}}$. The $n=0,1,2,...$ quantum number index the total wave-functions of distinct rotational bands of the same parity.

\subsection{Electromagnetic observables}

The electric transition probabilities can be written in the following compact form for $\lambda=0,1,2,3$:
\begin{equation}
B(E\lambda;L_{n}^{p}\rightarrow {L'}_{n'}^{p'})=t_{\lambda}\left(C^{L\,\lambda\,L'}_{0\,0\,0}\tilde{B}_{Lnp;L'n'p'}^{\lambda}I_{Lp;L'p'}^{\lambda}\right)^{2},\label{te}
\end{equation}
where $p=\pm$ is the state's parity. For $\lambda=0$, the relevant quantity is usually denoted as $\rho^{2}$. In the above expression, $C$ is the Clebsch-Gordan coefficient, $\tilde{B}$ are the integrals over the $\tilde{\beta}$ variable:
\begin{eqnarray}
\tilde{B}_{Lnp;L'n'p'}^{1,0}&=&\int_{0}^{\infty}\tilde{\beta}^{3}\psi_{Ln}^{p}(\tilde{\beta})\psi_{L'n'}^{p'}(\tilde{\beta})d\tilde{\beta},\\
\tilde{B}_{Lnp;L'n'p'}^{2,3}&=&\int_{0}^{\infty}\tilde{\beta}^{2}\psi_{Ln}^{p}(\tilde{\beta})\psi_{L'n'}^{p'}(\tilde{\beta})d\tilde{\beta},
\end{eqnarray}
while $I$ are the corresponding integrals over the angular variable $\phi$:
\begin{eqnarray}
I_{Lp;L'p'}^{0}&=&\int_{-\pi/2}^{\pi/2}\cos^{2}{\phi}\chi^{p}_{L}(\phi)\chi^{p'}_{L'}(\phi)d\phi,\\
I_{Lp;L'p'}^{1}&=&\int_{-\pi/2}^{\pi/2}\sin{2\phi}\chi^{p}_{L}(\phi)\chi^{p'}_{L'}(\phi)d\phi,\\
I_{Lp;L'p'}^{2}&=&\int_{-\pi/2}^{\pi/2}\cos{\phi}\chi^{p}_{L}(\phi)\chi^{p'}_{L'}(\phi)d\phi,\\
I_{Lp;L'p'}^{3}&=&\int_{-\pi/2}^{\pi/2}\sin{\phi}\chi^{p}_{L}(\phi)\chi^{p'}_{L'}(\phi)d\phi.
\end{eqnarray}
The integrated $\tilde{\beta}$ and $\phi$ functions are specific to the distinct transition operators \cite{Denisov1995,Bonatsos2015,Budaca2023}. The $t_{\lambda}$ constants are gathering the corresponding physical units and various normalization factors:
\begin{align}
&t_{0}=\left(\frac{3Z}{4\pi}\right)^{2}\frac{s^{4}}{16x^{3}}(x+1)^{4},\\
&t_{1}=\frac{3s^{4}}{4\pi}\left[\frac{9AZe^{3}}{56\sqrt{35}\pi}\left(\frac{1}{J}+\frac{15}{8QA^{1/3}}\right)\right]^{2}\frac{(x+1)^{2}}{16x},\\
&t_{2}=\left(\frac{3ZeR_{0}^{2}}{4\pi}\right)^{2}\frac{s^{2}}{8x^{2}}(x+1)^{3},\,\,\,t_{3}=R_{0}^{2}xt_{2}.
\end{align}
$R_{0}=1.2A^{1/3}$ fm is the nuclear radius, $Z$ and $A$ are the charge and mass numbers, $x=B_{2}/B_{3}$ and $s$ is a scaling factor relating the geometric quadrupole and octupole deformation parameters with their microscopic counterparts $\beta_{2,3}^{mic}=s\beta_{2,3}$. The expression for $t_{1}$ comes from the relation of the electric dipole transition operator with the polarized electric dipole moment in the first order approximation \cite{Denisov1995,Denisov2011}. The volume-symmetry energy $J=32.2$ MeV, and the effective surface stiffness $Q=28.72$ MeV are taken from the most recent Finite Range Droplet Model \cite{Moller2016}.

\section{Numerical application and discussions}
\label{Numerical_application_and_discussions}
\subsection{Model parameters}

\begin{figure*}
	\centering
	\includegraphics[width = 1.0\linewidth]{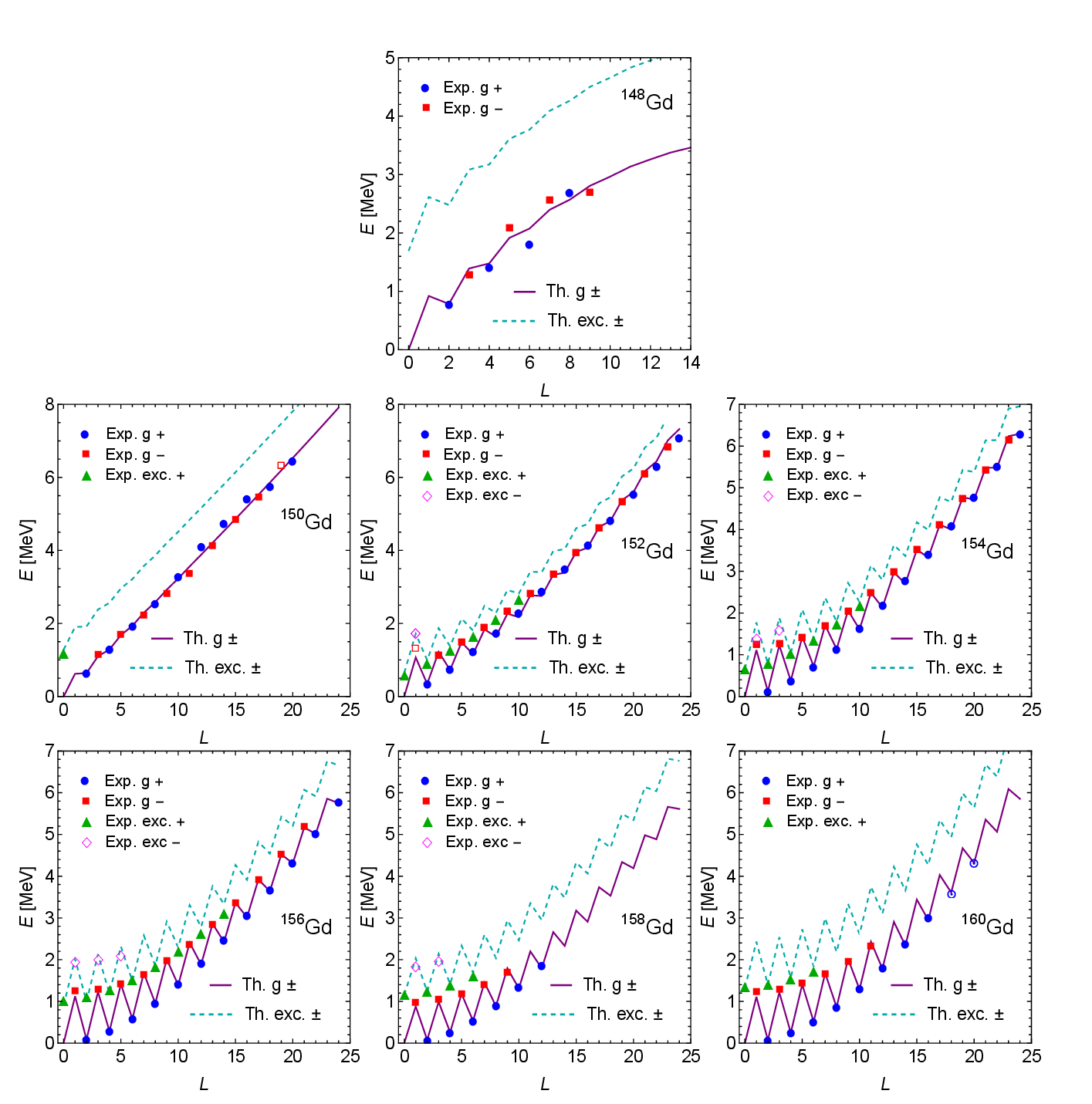}
 	\caption{Comparison between experimental and theoretical absolute energies for the positive- and negative-parity states of $^{148}$Gd \cite{Nica2014}, $^{150}$Gd \cite{Basu2013}, $^{152}$Gd \cite{Martin2013}, $^{154}$Gd \cite{Nica2025}, $^{156}$Gd \cite{Reich2012}, $^{158}$Gd \cite{Nica2017}, and $^{160}$Gd \cite{Nica2021,Hartley2021}. Open symbols indicate data points with uncertain band assignments but that conform to the theoretically predicted spin and parity. These levels are not included in the fitting procedure.}
	\label{Energy}
\end{figure*}

The independent model parameters $|\phi_{c}|$, $w_{0}$, and $c$ are determined by fitting the experimental energy of the yrast states with alternate parity and the positive parity states of the lowest excited band ($\beta$ band), with the energy expression (\ref{Etot}) for the Gd isotopes with $A=148-160$. The $\beta$ band states are considered as resulting from the $\tilde{\beta}$ excitation, with $n=1$. Numerical analysis showed that the $\phi$ mode of excitation provides higher excited levels for the considered nuclei. In what follows, the lower index of the states will represent their theoretical energy hierarchy. The fitting procedure is performed on level energies normalized to the excitation energy of the $2_{1}^{+}$ state. Absolute values of the theoretical energy levels are obtained by setting the energy scale such that the normalizing experimental value of the $2_{1}^{+}$ energy level is reproduced. The comparison between theoretical and experimental energy levels is shown in Fig.\ref{Energy}, while the obtained parameters are listed in Table \ref{tab1}. 

\setlength{\tabcolsep}{3.9pt}
\begin{table}[ht!]
\caption{Independent model parameters $|\phi_{c}|$, $w_{0}$, and $c$, as well as the scaling parameters $x$ and $s$ providing absolute electromagnetic transition probabilities.}
\label{tab1}
\vspace{0.2cm}
\begin{tabular}{clrr|rc}
\hline\hline\noalign{\smallskip}
Nucleus&$|\phi_{c}|$&$w_{0}\,\,\,$&$c\,\,\,\,\,\,\,\,$&$x=\frac{B_{2}}{B_{3}}$&$s$\\
\noalign{\smallskip}\hline\noalign{\smallskip}
$^{148}$Gd$_{84}$&20$^{\circ}$  & -7.330&-0.00593& 7.177&0.02454\\
$^{150}$Gd$_{86}$&17$^{\circ}$  & -7.772& 0.00129& 7.343&0.04677\\
$^{152}$Gd$_{88}$&75$^{\circ}$  & 36.806& 0.00949&29.219&0.09989\\
$^{154}$Gd$_{90}$&81.1$^{\circ}$&242.802& 0.00857& 6.068&0.17912\\
$^{156}$Gd$_{92}$&81.3$^{\circ}$&272.473& 0.00444& 4.760&0.11701\\
$^{158}$Gd$_{94}$&79.2$^{\circ}$&151.556& 0.00257& 0.235&0.06981\\
$^{160}$Gd$_{96}$&81.3$^{\circ}$&301.861& 0.00251& 1.388&0.14019\\
\noalign{\smallskip}\hline\hline
\end{tabular}
\end{table}

The model parameters $|\phi_{c}|$, $w_{0}$, and $c$, can be used further for predictions on transition probabilities. The ratios of transition probabilities with the same multipolarity exclude the scaling factors $t_{\lambda}$, and therefore constitute a good test of the theoretical formalism. Few such ratios are given in Table \ref{tab2}. We are also interested here in the absolute values of the transition rates, which can be directly confronted with experimental data. These are obtained by employing the scaling parameters $s$ and $x$, which are fixed by fitting all the available data on the $E0$, $E1$, $E2$, and $E3$ transition probabilities. By considering the lowest experimental errors as weighting factors, the difference in the range of values between different types of transition is eliminated. Moreover, for a consistent procedure for all considered nuclei, the residuals for each multipolarity are averaged on the available number of experimental data points. In the case of $E0$ transitions, one also included the $X(E0/E2)$ observable \cite{Rasmussen1960}. The resulting scaling parameters are listed in the last two columns of Table \ref{tab1}.

\subsection{Level energies}
The quality of the fit performed in this work for the positive and negative-parity levels of $^{148-160}$Gd is illustrated in Fig. \ref{Energy}. The experimental levels are taken from the Evaluated Nuclear Structure Data File (ENSDF) for each isotopic chain \cite{Nica2014,Basu2013,Martin2013,Nica2025,Nica2017,Nica2021}. For the lowest-mass isotopes in this work, $^{148}$Gd and $^{150}$Gd, one observes an almost alternating parity sequence of positive- and negative-parity states, starting as low as the $2^{+}_{1}$ and $3^{-}_{1}$ levels. This is a usual signature of reflection-asymmetric nuclei, although typical alternating parity bands start at slightly higher spin \cite{Butler1996}, with the negative-parity states shifted to higher energies. However, as mentioned in the introduction, the first excited level in $^{146}$Gd has $L^{\pi}=3^{-}$, and even if in heavier nuclei the $2^{+}_{1}$ state comes lower, it still retains an almost perfect alternating parity structure for $^{148}$Gd and $^{150}$Gd. Starting with $^{152}$Gd, the negative-parity states are shifted to higher energies, but they still form an alternating parity structure at higher spin. The parity splitting energy also increases with increasing mass. This interleaving pattern of positive- and negative-parity states suggests that the octupole deformation is seen in our calculations up to $N=88$, while for heavier isotopes only octupole vibrations are observed.

The angular momentum spectrum of the $^{148}$Gd nucleus is specific to the seniority scheme. The next nucleus, $^{150}$Gd, exhibits a linear dependence of the energy levels on spin. This is consistent with the vibrational character of near-spherical nuclear shapes. Starting with the critical point nucleus $^{154}$Gd ($N=90$), the rotational level sequences exhibit a clear parabolic behavior. Although the model is specifically tailored for describing the parity splitting mechanism, it is also capable of reproducing these distinct rotational features. This aspect is essential for the description of the two-fold shape-phase transition in the Gd isotopes. It would be interesting to test the predictions of this model for the heavier Gd isotopes, for which there is limited experimental information on the energy levels of high-spin states, especially for the negative-parity bands.

We note that the yrast sequence of negative-parity levels in each nucleus is reported as having a $K^{\pi}=1^{-}$ projection, except for the band belonging to $^{160}$Gd \cite{Nica2021} which has $K^{\pi}=0^{-}$. This is in contradiction with previous theoretical calculations, which considered the random-phase approximation framework, pointing out that in the rare-earth nuclei, the $K^{\pi}=0^{-}$ band comes lowest for the lightest nuclei in this region, while $K^{\pi}=1^{-}$ and $K^{\pi}=2^{-}$ bands share the lowest position in heavier mass nuclei \cite{Neergard1970}. In our calculations, all negative-parity states have $K^{\pi}=0^{-}$ for each nucleus. 

\begin{figure*}
	\centering
	\includegraphics[width = 1\linewidth]{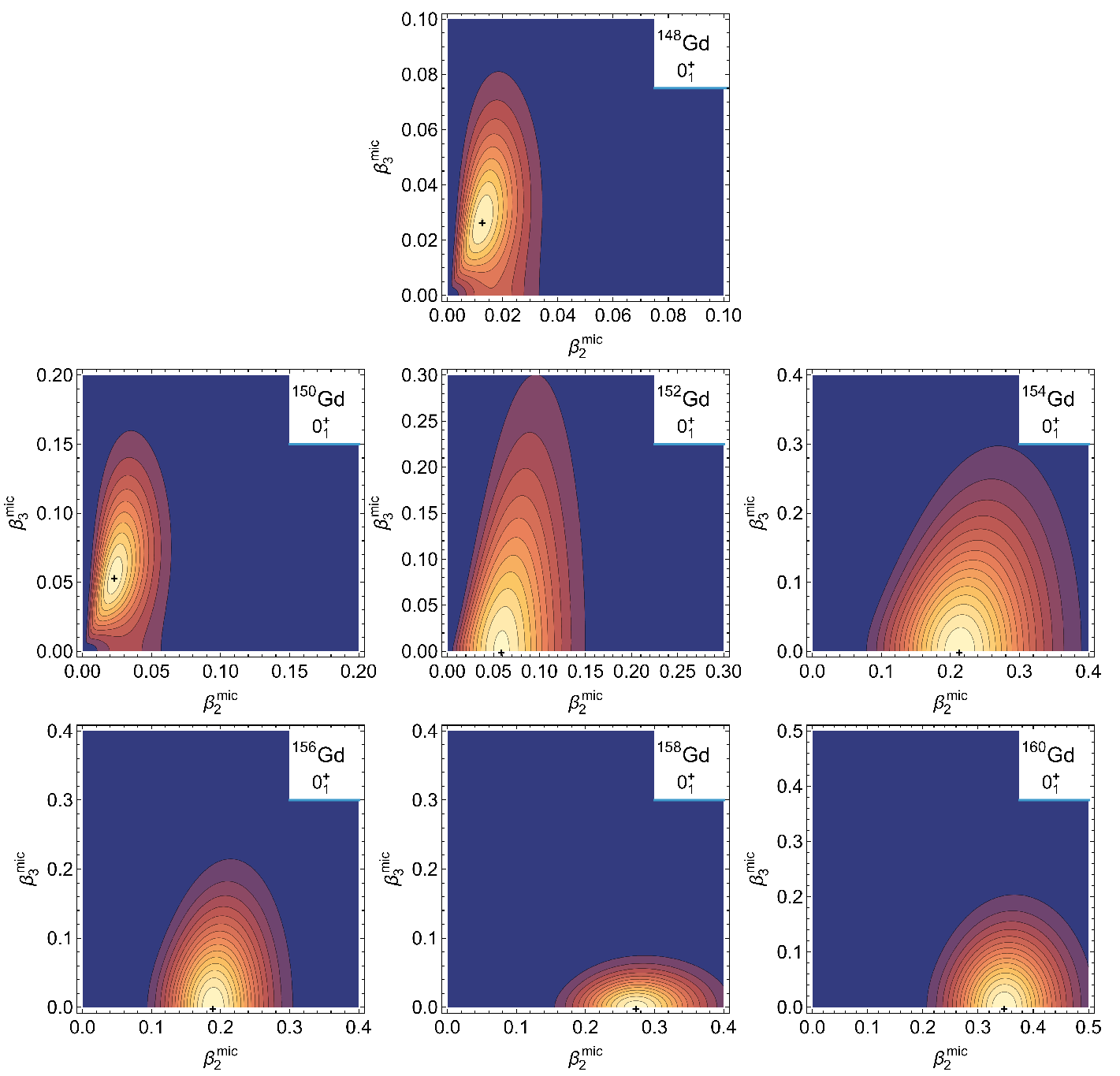}
 	\caption{Ground state deformation probability distribution for the considered nuclei, as a function of $\beta_{2}^{mic}$ and $\beta_{3}^{mic}$. The difference between consecutive contours is 0.05.}
	\label{Density}
\end{figure*}

\begin{figure}
	\centering
	\includegraphics[width = 1.0\linewidth]{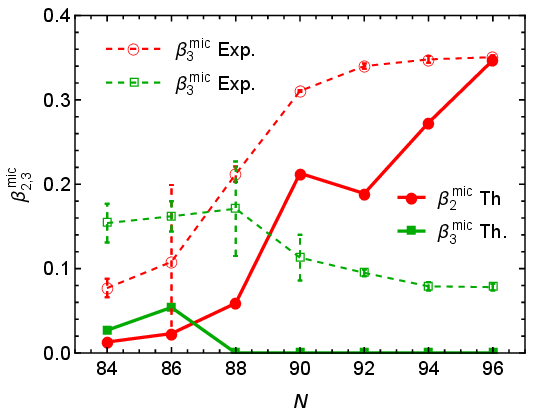}
 	\caption{Evolution of the theoretical and experimentally deduced \cite{Nudat,Kibedi2002} ground state equilibrium $\beta_{2}^{mic}$ and $\beta_{3}^{mic}$ deformations along the considered sequence of the $^{148-160}$Gd isotopes.}
	\label{beta23}
\end{figure}

\begin{figure}
	\centering
	\includegraphics[width = 1\linewidth]{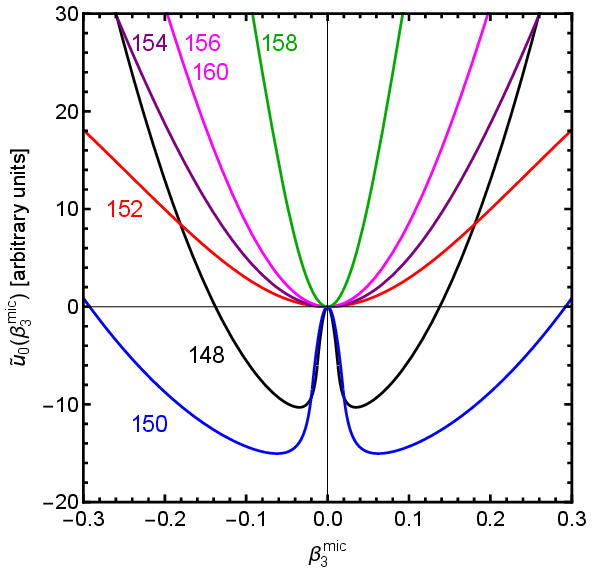}
 	\caption{Ground state potential $\tilde{u}_{0}(\phi)$ [Eq.(\ref{poti})] of the considered nuclei, as a function of $\beta_{3}^{mic}$.}
	\label{pot}
\end{figure}

The performance of the model is validated by matching the theoretical predictions of excited states with experimental data that have consistent spin and parity but uncertain band assignments. In this way, one can confirm the continuation of the negative- and positive-parity yrast bands in the $^{150}$Gd and, respectively, the $^{160}$Gd nucleus. Note that in the $^{152}$Gd nucleus, $E(1_{1}^{-})>E(3^{-}_{1})$, and therefore the $1_{1}^{-}$ energy level was not considered in the fitting procedure. Nevertheless, the predicted theoretical $1_{1}^{-}$ level energy is quite close to the experimental data point. One must also mention that the deformed $^{152-160}$Gd nuclei have excited negative parity states that conform very well with the theoretical negative parity partner of the $\beta$ band. The most suitable candidate states are shown in Fig. \ref{Energy}. The exception is the $^{160}$Gd nucleus, where there is a high density of matching states $1_{2}^{-}$ in the immediate vicinity of the theoretical predictions. In $^{152,154}$Gd and $^{158}$Gd, the corresponding experimental excited states have a $K^{\pi}=1^{-}$ assignment consistent with the reported yrast negative parity band. For the suggested experimental realization of the excited negative parity band in $^{156}$Gd, one does not have information on the projection of the $1_{2}^{-}$ state, while $3_{2}^{-}$ and $5_{2}^{-}$ levels have $K^{\pi}=2^{-}$ and respectively $K^{\pi}=4^{-}$.

The non-yrast positive-parity levels all have $K^{\pi}=0^{+}$ in the experiment and calculations, and are often identified with the presence of a $\beta$ vibration. While this identification, based on the original picture of a $\beta$ vibration proposed by Bohr-Mottelson, requires a multi-messenger approach involving different types of experimental information \cite{Garrett2001}, the presence of $\beta$-vibrational excitations was confirmed for the Gd isotopes in the case of $^{154}$Gd, $^{156}$Gd, and $^{158}$Gd \cite{Aprahamian2025}. The present calculations reproduce well the locations of the known excited levels and, as shown in a later section, the enhanced $E2$ strengths from these states support the $\beta$ vibration picture. This is especially important, since other theoretical models usually struggle to reproduce the correct rotational sequence of the $\beta$ bands. 

\begin{figure*}
	\centering
	\includegraphics[width = 0.9\linewidth]{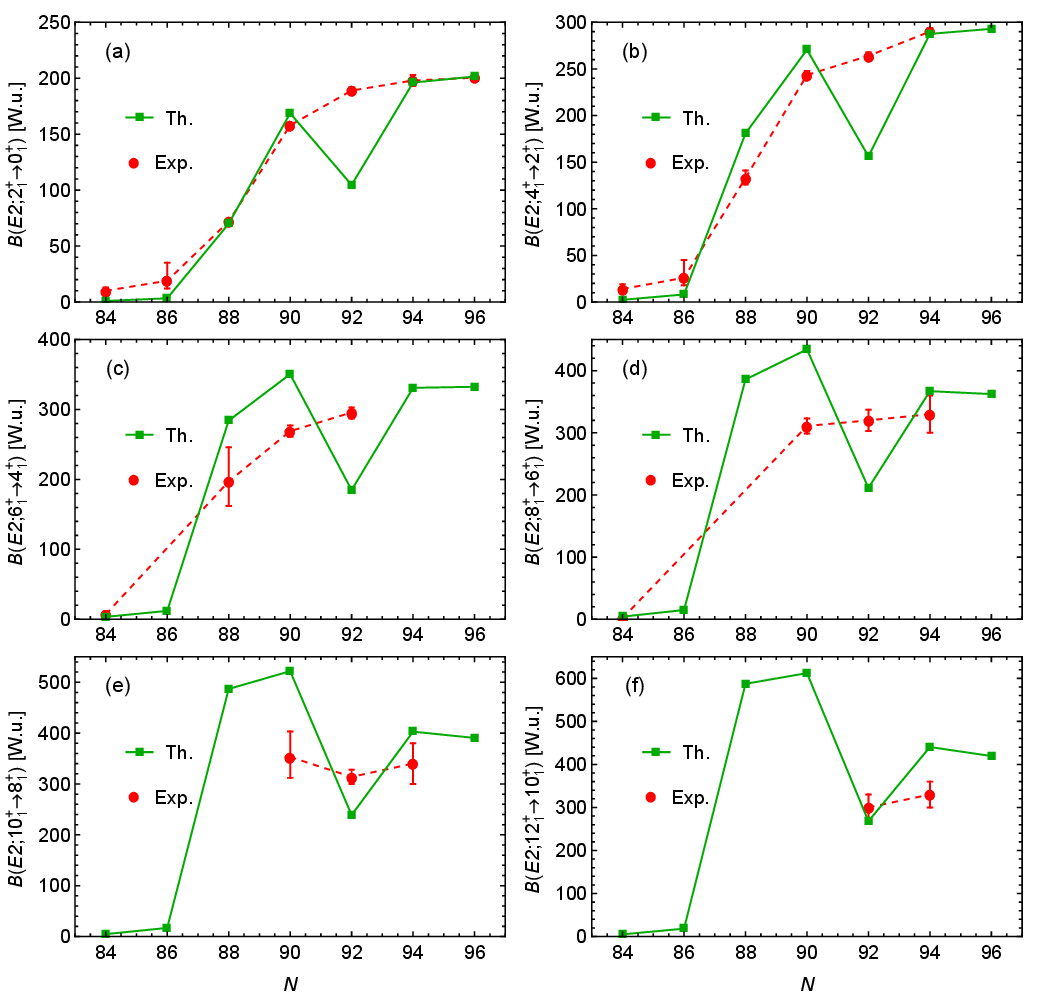}
 	\caption{Experimental \cite{Nica2014,Basu2013,Martin2013,Nica2025,Reich2012,Nica2017,Nica2021,Pascu2025,Knafla2023} and theoretical $E2$ transition rates between yrast positive parity states of the considered nuclei, as a function of the neutron number.}
	\label{E2in}
\end{figure*}

\begin{figure*}
	\centering
	\includegraphics[width = 0.9\linewidth]{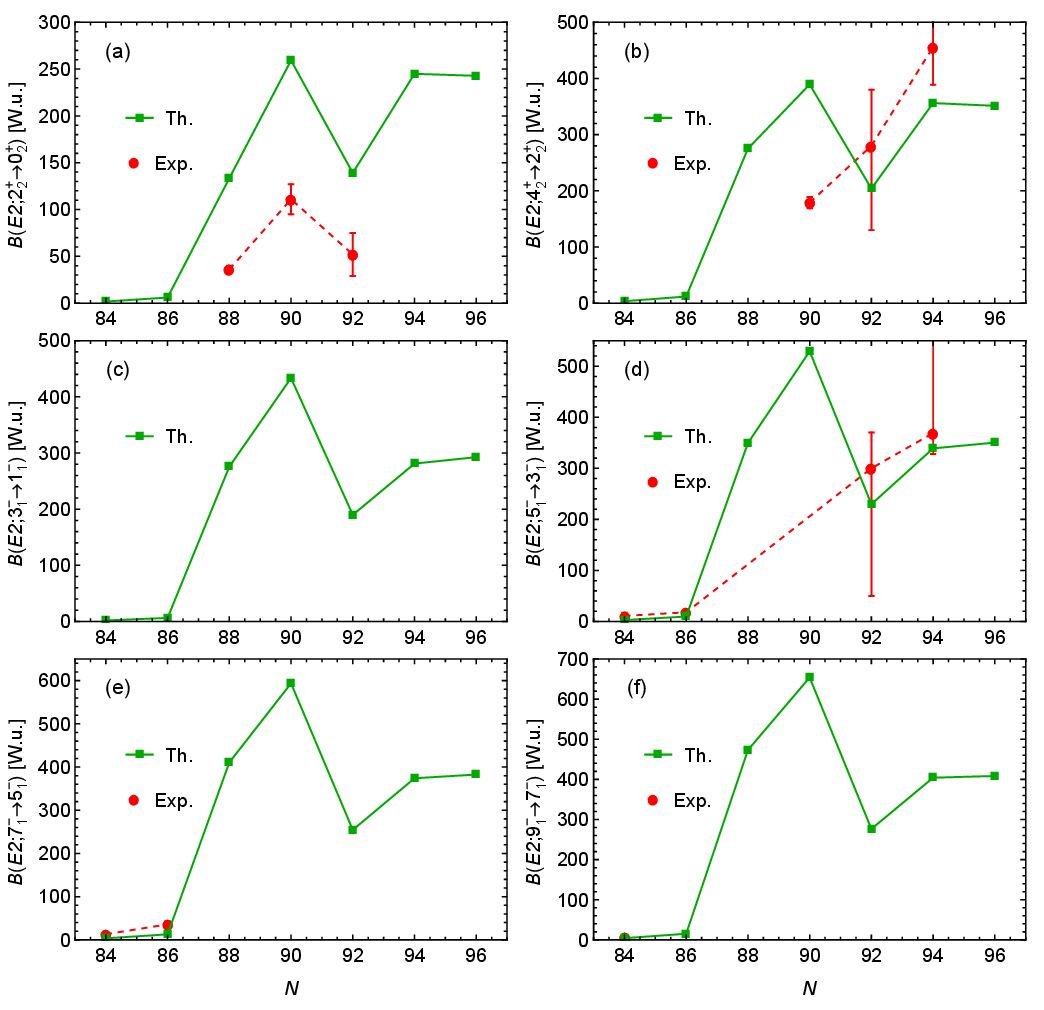}
 	\caption{Experimental \cite{Nica2014,Basu2013,Martin2013,Nica2025,Reich2012,Nica2017,Pascu2025} and theoretically predicted in-band $E2$ transition rates between excited states of positive parity (a,b) and of negative parity (c-f), as a function of the neutron number.}
	\label{E2exc}
\end{figure*}

\setlength{\tabcolsep}{3.9pt}
\begin{table*}[ht!]
\caption{Experimentally available ratios of $E0/E2$ \cite{Kibedi2005}, $E1$, $E2$, and $E3$  \cite{Nica2014,Martin2013,Nica2025,Reich2012,Nica2017,Pascu2025,Knafla2023} transition probabilities, for the considered nuclei, are compared with the model predictions. Only for the $E0/E2$ observable, the additional scale parameters $s$ and $x$ were necessary. The uncertainties are deduced through the propagation of errors. Transitions are considered from levels of higher energy.}
\label{tab2}
\vspace{0.2cm}
\begin{tabular}{ccccccccccccc}
\hline\hline\noalign{\smallskip}
&\multicolumn{2}{c}{$X(E0/E2)\times10^3$}&\multicolumn{2}{c}{$\displaystyle\frac{B(E1,1_{1}^{-}\rightarrow2_{1}^{+})}{B(E1,1_{1}^{-}\rightarrow0_{1}^{+})}$}&\multicolumn{2}{c}{$\displaystyle\frac{B(E1,3_{1}^{-}\leftrightarrow4_{1}^{+})}{B(E1,3_{1}^{-}\rightarrow2_{1}^{+})}$}&\multicolumn{2}{c}{$\displaystyle\frac{B(E1,5_{1}^{-}\leftrightarrow6_{1}^{+})}{B(E1,5_{1}^{-}\rightarrow4_{1}^{+})}$}&\multicolumn{2}{c}{$\displaystyle\frac{B(E2,4_{1}^{+}\rightarrow2_{1}^{+})}{B(E2,2_{1}^{+}\rightarrow0_{1}^{+})}$}&\multicolumn{2}{c}{$\displaystyle\frac{B(E3,5_{1}^{-}\rightarrow2_{1}^{+})}{B(E3,3_{1}^{-}\rightarrow0_{1}^{+})}$}\\
\noalign{\smallskip}\cline{2-13}\noalign{\smallskip}
Nucleus&Exp.&Th.&Exp.&Th.&Exp.&Th.&Exp.&Th.&Exp.&Th.&Exp.&Th.\\
\noalign{\smallskip}\hline\noalign{\smallskip}
$^{148}$Gd$_{84}$&       &0.3&                    &4.18&5.09(162)       &1.62& &1.66&1.40(65)&2.46& &2.95\\
$^{150}$Gd$_{86}$&17(4)  &1.0&                    &4.01&2.87$^{+208}_{-95}$&1.58& &1.38&1.37$^{+153}_{-66}$&2.45&1.18$^{+36}_{-30}$&2.88\\
$^{152}$Gd$_{88}$&12.2(5)&5.9&   1.53(5)          &5.83&     0.63(1)          &2.50& &1.86&1.85$^{+17}_{-16}$&2.58& &3.60\\
$^{154}$Gd$_{90}$&60(9)  &136&   1.26(4)          &2.42&     0.59(4)          &1.84& &1.68&1.54(3)&1.60& &1.68\\
$^{156}$Gd$_{92}$&191(15)&122&1.29(13)            &2.12&0.78(4)        &1.51&0.76(7)&1.41&1.40(3)&1.49& &1.54\\
$^{158}$Gd$_{94}$&       &258&0.98(7)             &2.06&0.89(2)        &1.42&0.79(7)&1.31&1.46(4)&1.47& &1.53\\
$^{160}$Gd$_{96}$&       &434&1.79(3)             &2.04&0.87(1)                &1.39& &1.28& &1.45& &1.52\\
\noalign{\smallskip}\hline\hline
\end{tabular}
\end{table*}

\subsection{Ground state deformation}
The scaling parameters $x$ and $s$ can be used to cast all results in the traditional deformation space. In Fig. \ref{Density}, we therefore present the ground state probability distribution for the $^{148-160}$Gd isotopes in the ($\beta_2^{mic}$, $\beta_3^{mic}$) plane. One can see that the quadrupole deformation increases relatively monotonically from small values around 0.01 in $^{148}$Gd, corresponding to spherical nuclei, to large values around 0.3-0.35 in $^{158}$Gd and $^{160}$Gd, indicative of axially quadrupole deformed shapes. An interesting jump is observed between $^{152}$Gd ($N = 88$) and $^{154}$Gd ($N=90$), where quadrupole deformation develops in an abrupt mode, similar to the predictions of the $X(5)$ critical point symmetry \cite{Iachello2001}. On the $\beta_3^{mic}$ axis, there are only two nuclei with an octupole deformation parameter different from zero, $^{148}$Gd and $^{150}$Gd, showing a well-developed octupole minimum around values of 0.03-0.05. Starting with $^{152}$Gd, although the minimum is predicted at zero, there is a clear spread in the probability distribution, indicative of an octupole soft mode. This effect is more pronounced in $^{152}$Gd, while $^{154}$Gd and $^{156}$Gd show the same effect, but to a lesser extent. 

Fig. \ref{beta23} illustrates the discussed evolution with neutron number of the equilibrium $\beta_2^{mic}$ and $\beta_3^{mic}$ deformation coordinates, corresponding to the maximum probability distribution of the ground state.  It is also compared with the experimental values extracted using the standard formalism of Bohr and Mottelson \cite{Bohr_Mottelson}. Although these experimental data are extracted from measured $B(E2;2_{1}^{+}\rightarrow0_{1}^{+})$ and $B(E3;3_{1}^{-}\rightarrow0_{1}^{+})$ values, they are still model dependent. Nevertheless, they provide a useful way to visualize the equilibrium deformation of a ground-state nucleus. The quadrupole deformation increases with the neutron number in both experiment and theory, with the calculated values systematically lower than the experimental ones. Additionally, a kink is observed at $N=92$ in the calculated values, whereas the experimental data exhibit a smoother evolution. However, the phase transition at $N=90$ is well reproduced, with deformation values jumping from about 0.05 at $N=88$ to values around 0.2 at $N=90$. The theoretical octupole deformation parameter is zero for all heavier Gd nuclei, except for $^{148}$Gd and $^{150}$Gd, where small values below 0.1 are obtained. In the experiment, the values are placed higher than the model data, with a maximum at $N=88$. Despite the quantitative differences, the overall experimental trends are well reproduced for both quadrupole and octupole deformations.

The inspection of the potential energy surfaces provides a complementary view of the structure of the nuclei in this study. The conclusions drawn above from the analysis of the ground state wave function are consistent with the shape of the corresponding potential $\tilde{u}_{0}(\phi)$ [see (Eq.\ref{poti})], responsible for the parity splitting mechanism. For a consistent correlation, the angular variable is expressed in terms of $\beta_{3}^{mic}$ as $\phi=\textrm{Arctan}[\beta_{3}^{mic}/(\langle\beta_{2}^{mic}\rangle\sqrt{x})]$, using the equilibrium ground state quadrupole deformation $\langle\beta_{2}^{mic}\rangle$ shown in Fig. \ref{beta23}. Thus, the non-zero octupole deformation of the two lightest nuclei is related to a double-well shape of the potential (see Fig. \ref{pot}). On the other hand, the heavier isotopes present a single minimum potential of variable shallowness, which is related to the vibrational character of their octupole deformation.

\subsection{Transition probabilities}
\subsubsection{Ratios of transition strengths}
\label{Ratios_of_transition_strengths}
We start the discussion of transition strengths with the presentation of transition probability ratios. These have the advantage that, except for the $E0/E2$ ratios, the additional scale parameters $s$ and $x$ cancel each other. The comparison of the calculated and experimental values for the $E0/E2$, $E1$, $E2$, and $E3$ ratios is summarized in Table \ref{tab2}, with the model reproducing the available experimental data reasonably well. The experimental $E1$ ratios involving two transitions starting from the same initial level are evaluated with the expression:

\begin{equation}
\frac{B(E1;J_i\rightarrow J_{f1})}{ B(E1;J_i\rightarrow J_{f2})}=\frac{I_{\gamma1}(E1)}{I_{\gamma2}(E1)}\frac{ E_{\gamma2}^{3}(E1)}{ E_{\gamma1}^{3}(E1)}
\end{equation}
where $I_{\gamma k}(E1)$ and $E_{\gamma k}(E1)$ ($k=1,2$) are the intensity and the energy of the $J_i\rightarrow J_{fk}$ transition, respectively. The only exceptions are the $3^{-}\rightarrow4^{+}$ and $5^{-}\rightarrow6^{+}$ transitions for $^{148}$Gd and $^{150}$Gd, for which the absolute values are used to calculate the ratios, as in these isotopes, the $3^{-}_{1}$ and $5^{-}_{1}$ states come lower in energy than the $4^{+}_{1}$ and $6^{+}_{1}$ levels, respectively. For the other multipolarities, as they start from different initial states, the ratio of absolute transition strengths is used. 

The model reproduces the experimental $X(E0/E2)$ ratios \cite{Kibedi2005} remarkably well, starting from low values in lighter nuclei and increasing gradually towards heavier mass isotopes. Although presented later in the discussion section, we only mention here that absolute 0$_{2}^{+}\rightarrow0_{1}^{+}$ transitions are also in good agreement with the calculations. Moreover, the model also reproduces other $E0$ transitions between levels with $L>0$ relatively well, as discussed later in section \ref{Absolute_transition_strengths}.  

\begin{figure}[!htbp]
	\centering
	\includegraphics[width = 0.99\linewidth]{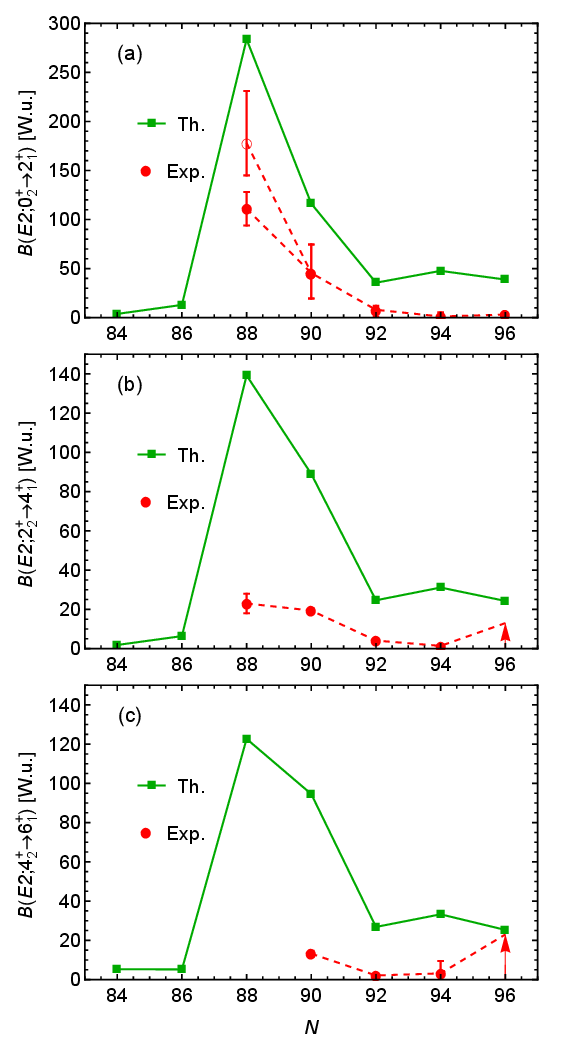}
 	\caption{Experimental \cite{Martin2013,Wiederhold2016,Reich2012,Nica2017,Nica2021,Pascu2025} and theoretically predicted inter-band $B(E2; L^{+}_{2}\rightarrow(L+2)^{+}_{1})$ transition probabilities between states of positive parity, as a function of the neutron number. The open symbol in panel (a) denotes an older data point \cite{Martin2013}, while the arrows indicate the upper experimental bounds.}
	\label{E2out1}
\end{figure}

\begin{figure*}
	\centering
	\includegraphics[width = 0.9\linewidth]{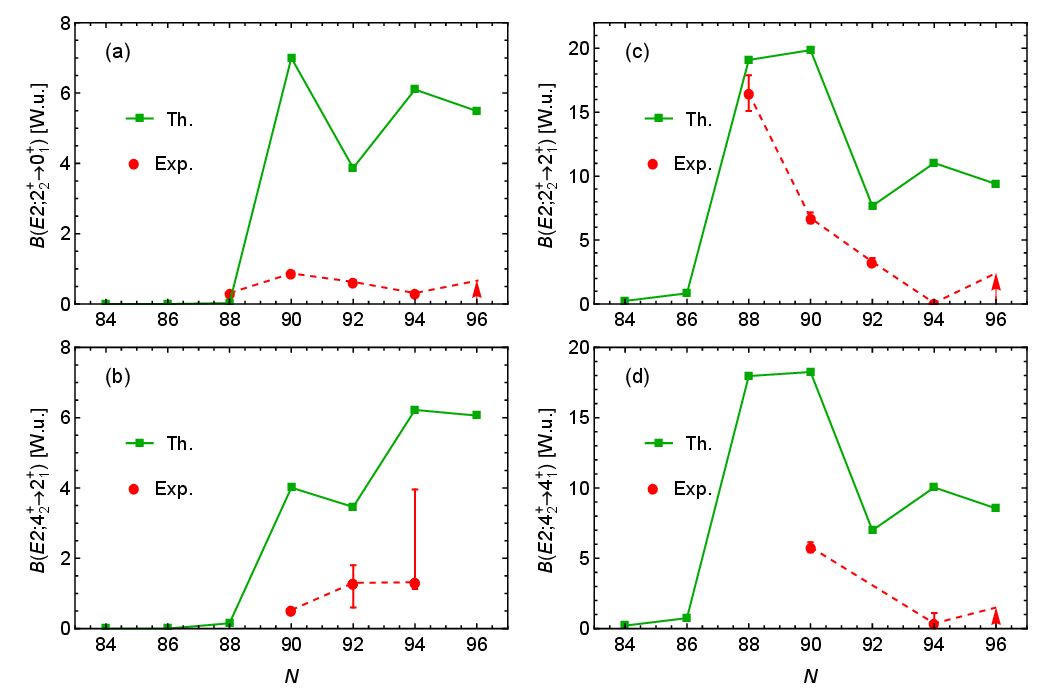}
 	\caption{Experimental \cite{Martin2013,Nica2025,Reich2012,Nica2017,Nica2021,Pascu2025,Lesher2015} and theoretically predicted inter-band $B(E2; L^{+}_{2}\rightarrow(L-2)^{+}_{1})$ (a,b) and $B(E2; L^{+}_{2}\rightarrow L^{+}_{1})$ (c,d) transition probabilities between states of positive parity, as a function of the neutron number. The arrows indicate upper experimental limits.}
	\label{E2out2}
\end{figure*}

\begin{figure*}
	\centering
	\includegraphics[width = 0.9\linewidth]{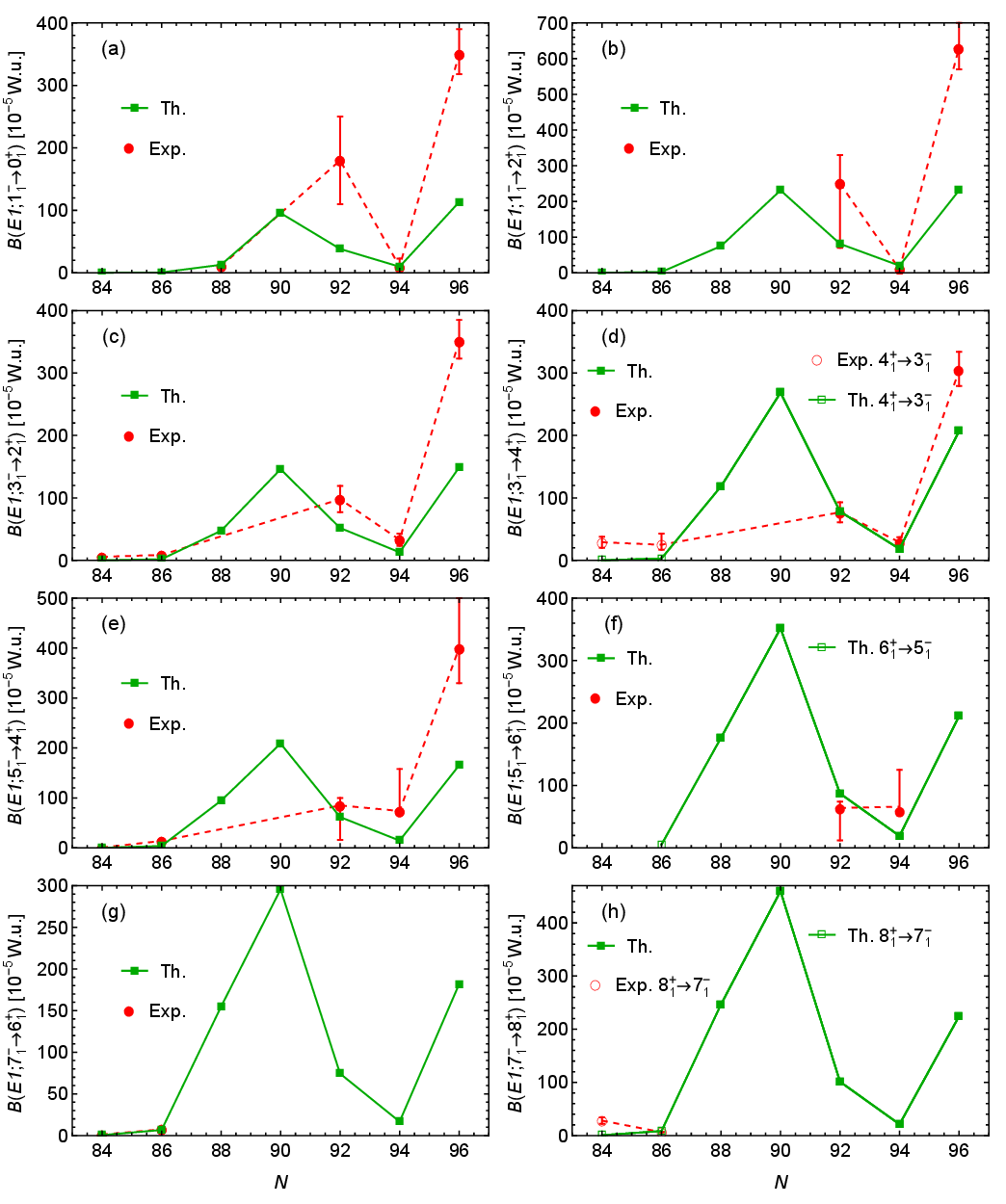}
 	\caption{Experimental \cite{Nica2014,Basu2013,Martin2013,Nica2025,Reich2012,Nica2017,Nica2021,Pascu2025} and theoretical $E1$ transition probabilities between yrast states, as a function of the neutron number.}
	\label{E1}
\end{figure*}

\begin{figure}
	\centering
	\includegraphics[width = 1.0\linewidth]{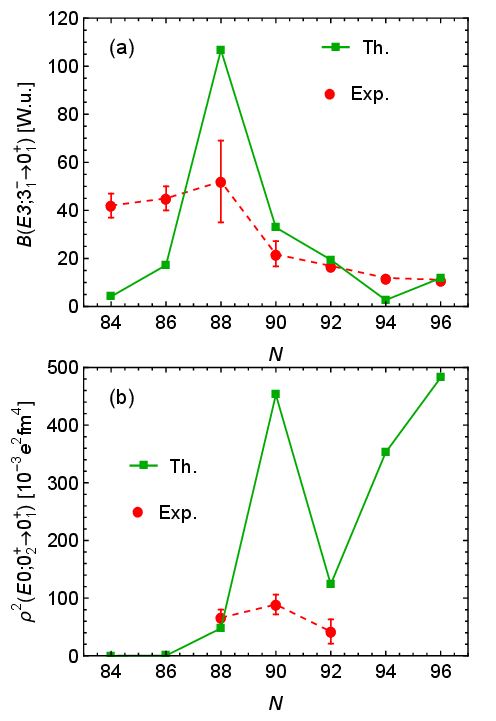}
 	\caption{Evolution of the theoretical and experimental $E3$ transition rates (a) \cite{Nica2014,Pascu2025,Kibedi2002,Wollersheim1977,McGowan1981}, and $E0$ transition strength (b) \cite{Martin2013,Kibedi2005,Reich2012}, in the considered sequence of the $^{148-160}$Gd nuclei.}
	\label{E3E0}
\end{figure}

As mentioned earlier, the yrast negative-parity states in Gd nuclei are considered to have the angular momentum projection $K^{\pi}=1^{-}$ in all nuclei, except for $^{160}$Gd, where $K^{\pi}=0^{-}$. According to the Alaga rule, the $B(E1;1^{-}_{1}\rightarrow2^{+}_{1})/B(E1;1^{-}_{1}\rightarrow0^{+}_{1})$ ratio is 2.0 for $K^{\pi}= 0^{-}$ and 0.5 for $K^{\pi}=1^{-}$. As seen in Table \ref{tab2}, most of the experimental values lie in between these limits, indicating a relatively strong band mixing. In our model, the $1^{-}_{1}$ states have $K^{\pi}=0^{-}$ and the ratios are calculated around the Alaga rule limit for this value, except for the spherical $^{148}$Gd and $^{150}$Gd isotopes, as well as for the transitional $^{152}$Gd, for which even higher values are found. The latter case supports the distinct nature of the measured $1_{1}^{-}$ state in $^{152}$Gd, and its consequent omission from the fitting procedure. The experimental $E1$ ratios from the first $3^{-}_{1}$ and $5^{-}_{1}$ states have values much closer to the Alaga rule for $K^{\pi}=1^{-}$ states (0.75 and 0.83, respectively), indicating less mixing with increasing spin. The calculated values, on the other hand, for the corresponding states are higher, but still in fair agreement with the experiment.

The vibrational and rotational limits for the $B(E2;4_{1}^{+}\rightarrow2_{1}^{+})/B(E2;(2_{1}^{+}\rightarrow0_{1}^{+})$ ratio are 2.0 and 1.43, respectively \cite{ Bohr_Mottelson}. The experimental values in Gd isotopes are closer to the rotational limit (see Table \ref{tab2}), even for the spherical nuclei at the beginning of the shell ($^{148}$Gd and $^{150}$Gd). However, for these nuclei, the uncertainties are relatively large, covering both limits. The calculated ratios start with higher values, close to the vibrational limit, and then decrease abruptly towards rotational values at $N=90$, where the phase transition occurs. Note that the weakly deformed $^{148-152}$Gd nuclei are expected to exhibit substantial mixing between the ground state band and the $\gamma$ band, which can account for the difference between theory and experiment. The $\gamma$ degree of freedom is, however, not considered in the present formalism, which is essentially based on the rigid $\gamma=0^{\circ}$ presumption. The impact of the deviations from axial symmetry on the classification of quadrupole-octupole excitations was previously studied in Ref. \cite{Bizzeti2004} with a more general instance of the Bohr Hamiltonian. Presently, there exists an incipient attempt to include the triaxiality in the quadrupole-octupole dynamics \cite{Xiang2025}, which might prove useful for the reproduction of the finer spectroscopic aspects. The agreement with experiment for the $E2$ transition ratios can be alternatively improved by considering a second-order contribution to the $E2$ operator \cite{Caprio2007}. As a matter of fact, the relative importance of the anharmonic term was shown to decrease with the neutron number, being especially significant for the same lightest Gd nuclei \cite{Raduta2012}.

Finally, the $E3$ ratios are only known experimentally for $^{150}$Gd. The theoretical calculations are in fair agreement with the experimental result, yielding values that decrease from vibrational to rotational nuclei. More experimental data would be desirable to test this trend.

\subsubsection{Absolute transition strengths}
\label{Absolute_transition_strengths}
Using the $s$ and $x$ parameters in Table \ref{tab1}, we have calculated absolute $E0$, $E1$, $E2$, and $E3$ transition strengths for each Gd isotope, for both yrast and non-yrast states. In Fig. \ref{E2in}, we present the comparison between the experimental and calculated $B(E2)$ values in the yrast positive parity band up to $12^{+}$. The agreement is, in general, good, with some exceptions. The calculations follow the increasing trend with neutron number, except at $N=92$, where a sharp drop is observed. This local minimum contradicts the data at low spins, but it seems to be supported in the case of the $B(E2;10^{+}_{1}\rightarrow 8_{1}^{+})$ observable. We also note that starting with the $4_{1}^{+}\rightarrow2_{1}^{+}$ transition, a sudden increase in the $E2$ strength is observed from $N=86$ to $N=88$. While this is relatively moderate and in agreement with the experimental values for the $4_{1}^{+}\rightarrow2_{1}^{+}$ transition, for higher spins in the yrast band, this effect becomes more pronounced. More experimental values are needed in this region to test this intriguing prediction. 

In what follows, we will discuss the theoretical predictions for other $E2$ transition probabilities, which were not considered in the fitting procedure for the scaling parameters $x$ and $s$. Thus, in Fig. \ref{E2exc} we present the experimental and calculated in-band $E2$ transition strengths, for both positive- ($2_{2}^{+}\rightarrow0_{2}^{+}$ and $4_{2}^{+}\rightarrow2_{2}^{+}$ transitions) and negative-parity states ($3_{1}^{-}\rightarrow1_{1}^{-}$, $5_{1}^{-}\rightarrow3_{1}^{-}$, $7_{1}^{-}\rightarrow5_{1}^{-}$, and $9_{1}^{-}\rightarrow7_{1}^{-}$ transitions). Only a handful of experimental data are available, especially for the negative-parity states, but they agree reasonably well with the calculations. The same decrease in $B(E2)$ values observed at $N=92$ in the yrast band is also seen for these transitions. However, for the $2_{2}^{+}\rightarrow0_{2}^{+}$ transition, this seems to agree with the measured values, which have a maximum at $N=90$ and a decrease towards a smaller value for $N=92$. 

A comparison of the calculated $E2$ transition strengths with the experimental data is shown in Figs. \ref{E2out1} and \ref{E2out2} for the inter-band transitions from the $\beta$ band to the ground state band. Fig. \ref{E2out1} presents the $L^{+}_{2}\rightarrow(L+2)^{+}_{1}$ type of transitions, while Fig. \ref{E2out2} shows the $L^{+}_{2}\rightarrow(L-2)^{+}_{1}$ [panels (a) and (b)] and $L^{+}_{2}\rightarrow L^{+}_{1}$ transitions [panels (c) and (d)]. As a general remark, these calculated $B(E2)$ values tend to overestimate the experimental absolute values, but, in general, retain a rather good description of the experimental trend. The $0_{2}^{+}\rightarrow2_{1}^{+}$ and $2_{2}^{+}\rightarrow4_{1}^{+}$ transitions form a maximum at $N=88$ and a decreasing pattern towards lower values is seen for heavier nuclei. The $4_{2}^{+}\rightarrow6_{1}^{+}$ experimental values stop at $N=90$, but the decrease seen towards the heavier nuclei resembles the same behavior as for the other two transitions mentioned above. In all these cases, the theory predicts a sharp decrease in $B(E2)$ values for lighter isotopes, attaining values of only a few W.u. for $N=84$ and $N=86$. More experimental data could confirm this pattern.

The decay of the negative-parity states is usually difficult to describe by any theoretical model. $E1$ decays are naturally hindered by the lack of available orbitals to produce these transitions, while $E3$ decays become strong when specific orbitals can be connected. Reproducing both of them simultaneously while predicting the correct magnitude of the quadrupole strength is a challenge for any theoretical model. In the current work, we present our results for the $E1$ and $E3$ transition strengths. 

We start the discussion with the $E1$ transitions, presenting in Fig. \ref{E1} the calculated and experimental $E1$ decays of the negative-parity states up to 7$^{-}$ in the $K^{\pi}=0^{-}$ band, showing for each level the decay to the lower [$L\rightarrow(L-1)$] and higher [$L\rightarrow(L+1)$] spin in the ground state band. The experimental $E1$ strengths in the Gd region are of the 10$^{-4}$ W.u. order, with higher values around 10$^{-3}$ W.u. in $^{160}$Gd ($N=96$). However, since they have such small values and can be influenced by small admixtures in the wave functions, the $E1$ strengths do not follow a clear pattern, with some values decreasing suddenly [e.g., in $^{158}$Gd ($N=94$)]. Nevertheless, the calculated values reproduce this behavior remarkably well, with a deep minimum at $N=94$ and an increase at $N=96$. For the lighter nuclei, there is not enough experimental information to draw a conclusion, especially the $B(E1)$ values in $^{154}$Gd ($N=90$) are missing, where the model predicts the strongest octupole collectivity. One must remark, that although the $1_{1}^{-}$ energy level in $^{152}$Gd and its $E1$ transition to $0_{1}^{+}$ are not considered in any of the fitting stages, the later is reproduced very well. 

In Fig. \ref{E3E0} (a), we show the evolution of the experimental and calculated octupole strength in the Gd isotopic chain. The experimental values start with relatively high values around 40 W.u. in $^{148}$Gd ($N=84$) and continue to increase for $^{150}$Gd. We note that the experimental point we use for $^{152}$Gd does not come from a direct measurement (mean lifetime and branching ratio, or Coulomb excitation experiments), but from inelastic scattering experiments \cite{Kibedi2002}. However, this procedure is model dependent, and although we included it in Fig. \ref{E3E0} panel (a), this value should be regarded as tentative. Nevertheless, its presence in our figure is important, as if this value is correct, it suggests the maximum octupole collectivity is reached for $^{152}$Gd ($N=88$). All the other values for heavier nuclei are much smaller, typically below 20 W.u., in the experiment and calculations. Although the absolute values are not always in good agreement, the increasing trend in lighter isotopes, the maximum at $N=88$, as well as the decreasing trend in heavier Gd nuclei, are all beautifully reproduced. In fact, to our knowledge, this is the first model to reproduce the experimental octupole trend in Gd isotopes correctly and to produce a peak in the $E3$ strength at $N=88$, one of the locations of the “octupole driving numbers” \cite{Cottle1990, Butler2016, Martinou2024}.

We conclude with the discussion of absolute transition strengths with $E0$ transitions. These are presented in Fig. \ref{E3E0} panel (b), where a comparison of the experimental and calculated values for the $0_{2}^{+}\rightarrow0_{1}^{+}$ transitions is shown. The current model overestimates these values, but it retains the main trend. For example, the maximum at $N=90$ is consistent with the critical $X(5)$ nature \cite{Bonnet2009} of the $^{154}$Gd nucleus. The model predicts higher values for heavier nuclei, which await experimental confirmation. As mentioned earlier, we also calculated the $E0$ strength between states with the same angular momentum. As an example, we give below the experimental values in $^{156}$Gd \cite{Reich2012} in comparison with the model calculations: $\rho^{2}(E0; 2_{2}^{+}\rightarrow2_{1}^{+})$ = 54(4) versus 127.7, and $\rho^{2}(E0; 4_{2}^{+}\rightarrow4_{1}^{+})$ = 50$^{+25}_{-16}$ versus 135.8, respectively.

\section{Conclusions}
\label{Conclusions}
Detailed calculations were performed in the present work for the even-even $^{148-160}$Gd nuclei using the quadrupole-octupole Bohr Hamiltonian. We pursued a phenomenological but more profound understanding of the quadrupole and octupole features of these nuclei, which were proposed in a recent experimental study to display the strongest octupole characteristics in the rare-earth region. The theoretical framework treats collective motion in even-even nuclei under axial symmetry, constructing a collective Hamiltonian with kinetic terms for quadrupole and octupole variables, along with a potential energy that accommodates multiple minima. This approach provides insight into the shape evolution along isotopic chains and within separate rotational bands. This schematic phenomenological model uses a small number of adjustable parameters, which were fitted to the available experimental data in the Gd isotopic chain. The results include a critical view of the level energies and the $E0$, $E1$, $E2$, and $E3$ transition strengths. The level energies are reproduced remarkably well, even in weakly deformed nuclei for which the model is not expected to capture detailed spectroscopic observables. The model generally overestimates absolute transition strengths, but the experimental trend is consistently reproduced across all transition multipolarities investigated in the present study. Among all the experimental transition strengths investigated in this work, the reproduction of the $E1$ transitions is particularly notable, with the calculated values generally lying very close to the experimental ones, a level of agreement rarely achieved by existing theoretical models. We also note that the evolution of the $B(E3)$ values observed in experiments on Gd isotopes, with a maximum around $N=88$, is qualitatively reproduced by this simple model. While more experimental values are needed to test these predictions, there is also an increasing need for more systematic theoretical studies of neighboring isotopic chains, such as Nd, Sm, and Dy, all of which are expected to display strong octupole collectivity. Finally, the present results demand a revisit of the heavy nuclei, which were previously considered as just a proof of concept for the basic ingredients of the model \cite{Budaca2022,Budaca2023}.

\section{Acknowledgments}
This work was supported by grants from the Ministry of Research, Innovation, and Digitization, CNCS - UEFISCDI, project number PN-IV-P1-PCE-2023-0273, within PNCDI IV, and project number PN-23-21-01-01/2023.

\bibliography{BibFile.bib}

@article{Denisov1995,
  title = {Collective states of even-even and odd nuclei with $\beta_{2},\,\beta_{3},\,...,\,\beta_{N}$ deformations},
  author = {Denisov, V. Yu. and Dzyublik, A. Ya.},
  journal = {Nuclear Physics A},
  volume = {589},
  pages = {17},
  year = {1995}
}

@article{Denisov2011,
  title = {Polarized electric dipole moment of well-deformed reflection asymmetric nuclei},
  author = {Denisov, V. Yu.},
  journal = {The European Physical Journal A},
  volume = {47},
  pages = {80},
  year = {2011}
}

@article{Bonatsos2005,
  title = {Analytic description of critical-point actinides in a transition from octupole deformation to octupole vibrations},
  author = {Bonatsos, D. and Lenis, D. and Minkov, N. and Petrellis, D. and Yotov, P.},
  journal = {Physical Review C},
  volume = {71},
  pages = {064309},
  year = {2005}
}

@article{Bonatsos2015,
  title = {Octupole deformation in light actinides within an analytic quadrupole octupole axially symmetric model with a {D}avidson potential},
  author = {Bonatsos, D. and Martinou, A. and Minkov, N. and Karampagia, S. and Petrellis, D.},
  journal = {Physical Review C},
  volume = {91},
  pages = {054315},
  year = {2015}
}

@article{Budaca2022,
  title = {Nuclear collective motion of heavy nuclei with axial quadrupole and octupole deformation},
  author = {Budaca, R. and Buganu, P. and Budaca, A. I.},
  journal = {Physical Review C},
  volume = {106},
  pages = {014311},
  year = {2022}
}

@article{Budaca2023,
  title = {Quadrupole-octupole shape and dynamics of $^{222}$\textsc{R}a},
  author = {Budaca, R. and Buganu, P. and Budaca, A. I.},
  journal = {The European Physical Journal A},
  volume = {59},
  pages = {242},
  year = {2023}
}

@article{Budaca2024,
  title = {Quadrupole-octupole collective excitations in medium mass nuclei},
  author = {Budaca, R. and Budaca, A. I. and Buganu, P.},
  journal = {Physica Scripta},
  volume = {99},
  pages = {035309},
  year = {2024}
}

@article{Moller2016,
  title = {Nuclear ground-state masses and deformations: \textsc{FRDM}(2012)},
  author = {M\"{o}ller, P. and Sierk, A. J. and Ichikawa, T. and Sagawa, H.},
  journal = {Atomic Data and Nuclear Data Tables},
  volume = {109-110},
  pages = {1},
  year = {2016}
}

@article{Walet1989,
  title = {The doubly-magic character of $^{146}$\textsc{G}d and its relation to $^{208}$\textsc{P}b},
  author = {Walet, N. R. and Stoop, P. and Glaudemans, P. W. M.},
  journal = {Zeitschrift f\"{u}r Physik A Atomic Nuclei},
  volume = {332},
  pages = {9--16},
  year = {1989}
}

@article{Kleinheinz1979,
  title = {Particle hole yrast states in $^{146}${G}d and $^{147}${G}d and the $\mathit{Z}=64$ shell closure},
  author = {Kleinheinz, P. and Broda, R. and Daly, P. J. and Lunardi, S. and Ogawa, M. and Blomqvist, J.},
  journal = {Zeitschrift f\" ur Physik A Atoms and Nuclei},
  volume = {290},
  issue = {3},
  pages = {279--295},
  year = {1979},
  url = {https://doi.org/10.1007/BF01408545}
}

@article{Blomqvist1983,
  title = {Atomic masses above $^{146}${G}d derived from a shell model analysis of high spin states},
  author = {Blomqvist, Jan and Kleinheinz, Peter and Daly, Patrick J.},
  journal = {Zeitschrift f\" ur Physik A Atoms and Nuclei},
  volume = {312},
  issue = {1},
  pages = {27--41},
  year = {1983},
  url = {https://doi.org/10.1007/BF01411658}
}

@article{Ahmad1985,
  title = {Nuclear spins, moments, and changes of the mean square charge radii of $^{140-153}${E}u},
  author = {Ahmad, S. A. and Klempt, W. and Ekström, C. and Neugart, R. and Wendt, K.},
  journal = {Zeitschrift f\" ur Physik A Atoms and Nuclei},
  volume = {321},
  issue = {1},
  pages = {35--45},
  year = {1985},
  url = {https://doi.org/10.1007/BF01411941}
}

@article{Cakirli2008,
  title = {Empirical signature for shape transitions mediated by sub-shell changes},
  author = {Cakirli, R. B. and Casten, R. F.},
  journal = {Phys. Rev. C},
  volume = {78},
  issue = {4},
  pages = {041301},
  numpages = {5},
  year = {2008},
  month = {Oct},
  publisher = {American Physical Society},
  doi = {10.1103/PhysRevC.78.041301},
  url = {https://link.aps.org/doi/10.1103/PhysRevC.78.041301}
}

@article{Iachello2001,
  title = {Analytic Description of Critical Point Nuclei in a Spherical-Axially Deformed Shape Phase Transition},
  author = {Iachello, F.},
  journal = {Phys. Rev. Lett.},
  volume = {87},
  issue = {5},
  pages = {052502},
  numpages = {4},
  year = {2001},
  month = {Jul},
  publisher = {American Physical Society},
  doi = {10.1103/PhysRevLett.87.052502},
  url = {https://link.aps.org/doi/10.1103/PhysRevLett.87.052502}
}

@article{Casten2001,
  title = {Empirical Realization of a Critical Point Description in Atomic Nuclei},
  author = {Casten, R. F. and Zamfir, N. V.},
  journal = {Phys. Rev. Lett.},
  volume = {87},
  issue = {5},
  pages = {052503},
  numpages = {4},
  year = {2001},
  month = {Jul},
  publisher = {American Physical Society},
  doi = {10.1103/PhysRevLett.87.052503},
  url = {https://link.aps.org/doi/10.1103/PhysRevLett.87.052503}
}

@article{Krucken2002,
  title = {$\mathit{B}(\mathit{E}2)$ Values in $^{150}$\textsc{N}d and the Critical Point Symmetry $\mathit{X}(5)$},
  author = {Kr\"ucken, R. and Albanna, B. and Bialik, C. and Casten, R. F. and Cooper, J. R. and Dewald, A. and Zamfir, N. V. and Barton, C. J. and Beausang, C. W. and Caprio, M. A. and Hecht, A. A. and Klug, T. and Novak, J. R. and Pietralla, N. and von Brentano, P.},
  journal = {Phys. Rev. Lett.},
  volume = {88},
  issue = {23},
  pages = {232501},
  numpages = {4},
  year = {2002},
  month = {May},
  publisher = {American Physical Society},
  doi = {10.1103/PhysRevLett.88.232501},
  url = {https://link.aps.org/doi/10.1103/PhysRevLett.88.232501}
}

@article{Tonev2004,
  title = {Transition probabilities in $^{154}$\textsc{G}d: Evidence for {X}(5) critical point symmetry},
  author = {Tonev, D. and Dewald, A. and Klug, T. and Petkov, P. and Jolie, J. and Fitzler, A. and M\"oller, O. and Heinze, S. and von Brentano, P. and Casten, R. F.},
  journal = {Phys. Rev. C},
  volume = {69},
  issue = {3},
  pages = {034334},
  numpages = {6},
  year = {2004},
  month = {Mar},
  publisher = {American Physical Society},
  doi = {10.1103/PhysRevC.69.034334},
  url = {https://link.aps.org/doi/10.1103/PhysRevC.69.034334}
}

@article{Sheng2005,
author = {Sheng, Z.-Q. and Guo, J.-Y.},
title = {SYSTEMATIC ANALYSIS OF CRITICAL POINT NUCLEI IN THE RARE-EARTH REGION WITH RELATIVISTIC MEAN FIELD THEORY},
journal = {Modern Physics Letters A},
volume = {20},
number = {35},
pages = {2711-2721},
year = {2005},
doi = {10.1142/S0217732305017883}
}

@article{Fossion2006,
  title = {$\mathit{E}(5)$, $\mathit{X}(5)$, and prolate to oblate shape phase transitions in relativistic \textsc{H}artree-\textsc{B}ogoliubov theory},
  author = {Fossion, R. and Bonatsos, Dennis and Lalazissis, G. A.},
  journal = {Phys. Rev. C},
  volume = {73},
  issue = {4},
  pages = {044310},
  numpages = {10},
  year = {2006},
  month = {Apr},
  publisher = {American Physical Society},
  doi = {10.1103/PhysRevC.73.044310},
  url = {https://link.aps.org/doi/10.1103/PhysRevC.73.044310}
}

@article{Shi2018,
  title = {Low-lying states in even \textsc{G}d isotopes studied with five-dimensional collective \textsc{H}amiltonian based on covariant density functional theory},
  author = {Shi, Z. and Chen, Q. B. and Zhang, S. Q.},
  journal = {The European Physical Journal A},
  volume = {54},
  pages = {53},
  year = {2018}
}

@article{Naz2018,
title = {Microscopic description of structural evolution in {P}d, {X}e, {B}a, {N}d, {S}m, {G}d and {D}y isotopes},
journal = {Nuclear Physics A},
volume = {979},
pages = {1-20},
year = {2018},
issn = {0375-9474},
doi = {https://doi.org/10.1016/j.nuclphysa.2018.09.001},
url = {https://www.sciencedirect.com/science/article/pii/S0375947418301830},
author = {Tabassum Naz and G.H. Bhat and S. Jehangir and Shakeb Ahmad and J.A. Sheikh}
}

@article{Quan2018,
  title = {Nuclear quantum shape-phase transitions in odd-mass systems},
  author = {Quan, S. and Li, Z. P. and Vretenar, D. and Meng, J.},
  journal = {Phys. Rev. C},
  volume = {97},
  issue = {3},
  pages = {031301},
  numpages = {6},
  year = {2018},
  month = {Mar},
  publisher = {American Physical Society},
  doi = {10.1103/PhysRevC.97.031301},
  url = {https://link.aps.org/doi/10.1103/PhysRevC.97.031301}
}

@article{Guzman2007,
  title = {$\mathit{E}$(5) and $\mathit{X}$(5) shape phase transitions within a \textsc{S}kyrme-\textsc{H}artree-\textsc{F}ock + \textsc{BCS} approach},
  author = {Rodr\'{\i}guez-Guzm\'an, R. and Sarriguren, P.},
  journal = {Phys. Rev. C},
  volume = {76},
  issue = {6},
  pages = {064303},
  numpages = {8},
  year = {2007},
  month = {Dec},
  publisher = {American Physical Society},
  doi = {10.1103/PhysRevC.76.064303},
  url = {https://link.aps.org/doi/10.1103/PhysRevC.76.064303}
}

@article{Nomura2019,
  title = {Two-neutron transfer reactions and shape phase transitions in the microscopically formulated interacting boson model},
  author = {Nomura, K. and Zhang, Y.},
  journal = {Phys. Rev. C},
  volume = {99},
  issue = {2},
  pages = {024324},
  numpages = {11},
  year = {2019},
  month = {Feb},
  publisher = {American Physical Society},
  doi = {10.1103/PhysRevC.99.024324},
  url = {https://link.aps.org/doi/10.1103/PhysRevC.99.024324}
}

@article{Rodriguez2009,
    author = {Rodr\'{i}guez, Tomás R. and Egido, J. L.},
    title = {Study of shape transitions in $\mathit{N}\simeq90$ isotopes with beyond mean field calculations},
    journal = {AIP Conference Proceedings},
    volume = {1090},
    number = {1},
    pages = {419-423},
    year = {2009},
    month = {01},
    issn = {0094-243X},
    doi = {10.1063/1.3087058},
    url = {https://doi.org/10.1063/1.3087058}
}

@article{Pascu2025,
  title = {Increasing Octupole Collectivity across the $\mathit{Z}=64$ Isotopic Chain: $\mathit{B}(\mathit{E}3)$ Values in $^{150}$\mathsc{G}d},
  author = {Pascu, S. and Y\"uksel, E. and Abhishek and Stevenson, P. and Bhat, G. H. and Mao, R. N. and Nomura, K. and Costache, C. and Li, Z. P. and M\ifmmode \u{a}\else \u{a}\fi{}rginean, N. and Mihai, C. and Naz, T. and Podoly\'ak, Zs. and Regan, P. H. and Turturic\ifmmode \u{a}\else \u{a}\fi{}, A. E. and Borcea, R. and Boromiza, M. and Bucurescu, D. and C\ifmmode \u{a}\else \u{a}\fi{}linescu, S. and Clisu, C. and Coman, A. and Dinescu, I. and Doshi, S. and Filipescu, D. and Florea, N. M. and Gandhi, A. and Gheorghe, I. and Ionescu, A. and Lic\ifmmode \u{a}\else \u{a}\fi{}, R. and M\ifmmode \u{a}\else \u{a}\fi{}rginean, R. and Mihai, R. E. and Mitu, A. and Nazir, N. and Negret, A. and Ni\ifmmode \mbox{\c{t}}\else \c{t}\fi{}\ifmmode \u{a}\else \u{a}\fi{}, C. R. and O'Sullivan, E. B. and Petrone, C. and Poulton, S. E. and Sheikh, J. A. and Singh, H. K. and Stan, L. and Toma, S. and Turturic\ifmmode \u{a}\else \u{a}\fi{}, G. and Ujeniuc, S.},
  journal = {Phys. Rev. Lett.},
  volume = {134},
  issue = {9},
  pages = {092501},
  numpages = {7},
  year = {2025},
  month = {Mar},
  publisher = {American Physical Society},
  doi = {10.1103/PhysRevLett.134.092501},
  url = {https://link.aps.org/doi/10.1103/PhysRevLett.134.092501}
}

@article{Bucher2016,
  title = {Direct Evidence of Octupole Deformation in Neutron-Rich $^{144}\mathrm{Ba}$},
  author = {Bucher, B. and Zhu, S. and Wu, C. Y. and Janssens, R. V. F. and Cline, D. and Hayes, A. B. and Albers, M. and Ayangeakaa, A. D. and Butler, P. A. and Campbell, C. M. and Carpenter, M. P. and Chiara, C. J. and Clark, J. A. and Crawford, H. L. and Cromaz, M. and David, H. M. and Dickerson, C. and Gregor, E. T. and Harker, J. and Hoffman, C. R. and Kay, B. P. and Kondev, F. G. and Korichi, A. and Lauritsen, T. and Macchiavelli, A. O. and Pardo, R. C. and Richard, A. and Riley, M. A. and Savard, G. and Scheck, M. and Seweryniak, D. and Smith, M. K. and Vondrasek, R. and Wiens, A.},
  journal = {Phys. Rev. Lett.},
  volume = {116},
  issue = {11},
  pages = {112503},
  numpages = {5},
  year = {2016},
  month = {Mar},
  publisher = {American Physical Society},
  doi = {10.1103/PhysRevLett.116.112503},
  url = {https://link.aps.org/doi/10.1103/PhysRevLett.116.112503}
}

@article{Tsunoda2023,
  title = {Shape transition of \textsc{N}d and \textsc{S}m isotopes and the neutrinoless double-$\ensuremath{\beta}$-decay nuclear matrix element of $^{150}$\textsc{N}d},
  author = {Tsunoda, Yusuke and Shimizu, Noritaka and Otsuka, Takaharu},
  journal = {Phys. Rev. C},
  volume = {108},
  issue = {2},
  pages = {L021302},
  numpages = {6},
  year = {2023},
  month = {Aug},
  publisher = {American Physical Society},
  doi = {10.1103/PhysRevC.108.L021302},
  url = {https://link.aps.org/doi/10.1103/PhysRevC.108.L021302}
}

@article{Guzman2023,
  title = {Beyond-mean-field description of octupolarity in dysprosium isotopes with the {G}ogny-{D1M} energy density functional},
  author = {Rodr\'{\i}guez-Guzm\'an, R. and Robledo, L. M.},
  journal = {Phys. Rev. C},
  volume = {108},
  issue = {2},
  pages = {024301},
  numpages = {10},
  year = {2023},
  month = {Aug},
  publisher = {American Physical Society},
  doi = {10.1103/PhysRevC.108.024301},
  url = {https://link.aps.org/doi/10.1103/PhysRevC.108.024301}
}

@article{Reich2012,
title = {Nuclear \textsc{D}ata \textsc{S}heets for $\mathit{A}=156$},
journal = {Nuclear Data Sheets},
volume = {113},
number = {11},
pages = {2537-2840},
year = {2012},
issn = {0090-3752},
doi = {https://doi.org/10.1016/j.nds.2012.10.003},
url = {https://www.sciencedirect.com/science/article/pii/S0090375212000798},
author = {C.W. Reich}
}

@article{Nica2014,
title = {Nuclear \textsc{D}ata \textsc{S}heets for $\mathit{A}=148$},
journal = {Nuclear Data Sheets},
volume = {117},
pages = {1},
year = {2014},
doi = {https://doi.org/10.1016/j.nds.2014.02.001},
author = {N. Nica},
}

@article{Basu2013,
title = {Nuclear \textsc{D}ata \textsc{S}heets for $\mathit{A}=150$},
journal = {Nuclear Data Sheets},
volume = {114},
pages = {435},
year = {2013},
doi = {https://doi.org/10.1016/j.nds.2013.04.001},
author = {Basu, S. K. and Sonzogni, A. A.},
}

@article{Martin2013,
title = {Nuclear \textsc{D}ata \textsc{S}heets for $\mathit{A}=152$},
journal = {Nuclear Data Sheets},
volume = {114},
pages = {1497},
year = {2013},
doi = {https://doi.org/10.1016/j.nds.2013.11.001},
author = {Martin, M. J.},
}

@article{Nica2025,
title = {Nuclear \textsc{D}ata \textsc{S}heets for $\mathit{A}=154$},
journal = {Nuclear Data Sheets},
volume = {200},
pages = {2},
year = {2025},
doi = {https://doi.org/10.1016/j.nds.2025.02.002},
author = {N. Nica},
}

@article{Nica2017,
title = {Nuclear \textsc{D}ata \textsc{S}heets for $\mathit{A}=158$},
journal = {Nuclear Data Sheets},
volume = {141},
pages = {1-326},
year = {2017},
issn = {0090-3752},
doi = {https://doi.org/10.1016/j.nds.2017.03.001},
url = {https://www.sciencedirect.com/science/article/pii/S009037521730025X},
author = {N. Nica}
}

@article{Nica2021,
title = {Nuclear \textsc{D}ata \textsc{S}heets for $\mathit{A}=160$},
journal = {Nuclear Data Sheets},
volume = {176},
pages = {1-428},
year = {2021},
issn = {0090-3752},
doi = {https://doi.org/10.1016/j.nds.2021.08.001},
url = {https://www.sciencedirect.com/science/article/pii/S0090375221000442},
author = {N. Nica}
}

@article{Hartley2021,
title = {Possible quenching of static neutron pairing near the $\mathit{N}=98$ deformed shell gap: \textsc{R}otational structures in $^{160,161}$\textsc{G}d},
journal = {Physical Review C},
volume = {103},
pages = {034322 },
year = {2021},
doi = {https://doi.org/10.1103/PhysRevC.103.034322},
author = {Hartley, D. J. and Villafana, K. and Kondev, F. G. and Riley, M. A. and Janssens, R. V. F. and Auranen, K. and Ayangeakaa, A. D. and Baron, J. S. and Boston, A. J. and Carpenter, M. P. and Clark, J. A. and Greene, J. P. and Heery, J. and Hoffman, C. R. and Jackson, P. and Lauritsen, T. and Li, J. and Little, D. and Paul, E. S. and Savard, G. and Seweryniak, D. and Simpson, J. and Stolze, S. and Wilson, G. L. and Wu, J. and Zhu, S. and Frauendorf, S}
}

@article{Kibedi2005,
title = {Electric monopole transitions between $0^{+}$ states for nuclei throughout the periodic table},
journal = {Atomic Data and Nuclear Data Tables},
volume = {89},
pages = {77},
year = {2005},
doi = {https://doi:10.1016/j.adt.2004.11.002},
author = {Kib\'{e}di, T and Spear, R. H.}
}

@article{Wiederhold2016,
  title = {Fast-timing lifetime measurement of $^{152}\mathrm{Gd}$},
  author = {Wiederhold, J. and Kern, R. and Lizarazo, C. and Pietralla, N. and Werner, V. and Jolos, R. V. and Bucurescu, D. and Florea, N. and Ghita, D. and Glodariu, T. and Lica, R. and Marginean, N. and Marginean, R. and Mihai, C. and Mihai, R. and Mitu, I. O. and Negret, A. and Nita, C. and Olacel, A. and Pascu, S. and Stroe, L. and Toma, S. and Turturica, A.},
  journal = {Phys. Rev. C},
  volume = {94},
  issue = {4},
  pages = {044302},
  numpages = {7},
  year = {2016},
  month = {Oct},
  publisher = {American Physical Society},
  doi = {10.1103/PhysRevC.94.044302},
  url = {https://link.aps.org/doi/10.1103/PhysRevC.94.044302}
}

@article{Wollersheim1977,
title = {$\mathit{E}2$ and $\mathit{E}3$ transition strengths in some rare-earth nuclei},
journal = {Zeitschrift f\"{u}r Physik A Atoms and Nuclei},
volume = {280},
pages = {277},
year = {1977},
doi = {https://doi.org/10.1007/BF01434354},
author = {Wollersheim, H. J. and Elze, Th. W.}
}

@article{McGowan1981,
title = {Reduced $\mathit{M}1$, $\mathit{E}1$, $\mathit{E}2$, and $\mathit{E⁢}3$ transition probabilities for transitions in $^{1⁢5⁢6-1⁢6⁢0}$\textsc{G}d and $^{1⁢6⁢0-1⁢6⁢4}$\textsc{D}y},
journal = {Physical Review C},
volume = {23},
pages = {1926},
year = {1984},
doi = {https://doi.org/10.1103/PhysRevC.23.1926},
author = {McGowan, F. K. and Milner, W. T.},
}

@article{Rasmussen1960,
title = {Theory of $\mathit{E}0$ transitions of spheroidal nuclei},
journal = {Nuclear Physics A},
volume = {19},
pages = {85},
year = {1960},
doi = {https://doi.org/10.1016/0029-5582(60)90221-2},
author = {Rasmussen, J. O.},
}

@article{Knafla2023,
title = {Improving fast-timing time-walk calibration standards: \textsc{L}ifetime measurement of the $2_{1}^{+}$ state in $^{152}$\textsc{G}d},
journal = {Nuclear Instruments and Methods in Physics Research Section A: Accelerators, Spectrometers, Detectors and Associated Equipment},
volume = {1052},
pages = {168279},
year = {2023},
doi = {https://doi.org/10.1016/j.nima.2023.168279},
author = {Knafla, L. and Harter, A. and Ley, M. and Esmaylzadeh, A. and R\'{e}gis, J.-M. and Bittner, D. and Blazhev, A. and von Spee, F. and Jolie, J.},
}

@misc{Xiang2025,
      title={Microscopic triaxial quadrupole-octupole collective \textsc{H}amiltonian for low-energy nuclear excitations}, 
      author={J. Xiang and J. Zhao and Z. P. Li and D. Vretenar},
      year={2025},
      eprint={2510.16453},
      archivePrefix={arXiv},
      primaryClass={nucl-th},
      url={https://arxiv.org/abs/2510.16453}, 
}

@article{Caprio2007,
title = {Analytic descriptions for transitional nuclei near the critical point},
journal = {Nuclear Physics A},
volume = {781},
pages = {26},
year = {2007},
doi = {https://doi.org/10.1016/j.nuclphysa.2006.10.032},
author = {Caprio, M. A. and Iachello, F},
}

@article{Raduta2012,
title = {Analytical description of the coherent state model for near vibrational and well deformed nuclei},
journal = {Annals of Physics (New York)},
volume = {325},
pages = {671},
year = {2012},
doi = {https://doi.org/10.1016/j.aop.2011.10.004},
author = {Raduta, A. A. and Budaca, R. and Faessler, A.},
}

@article{Bonnet2009,
title = {$\mathit{E}0$ transition strengths from $\mathit{X}(5)$ to the rigid rotor},
journal = {Physical Review C},
volume = {79},
pages = {034307},
year = {2009},
doi = {https://doi.org/10.1103/PhysRevC.79.034307},
author = {Bonnet, J. and Krugmann, A. and Beller, J. and Pietralla, N and Jolos, R. V.},
}

@article{Lesher2015,
title = {Collectivity of $0^{+}$ states in $^{160}$\textsc{G}d},
journal = {Physical Review C},
volume = {91},
pages = {054317},
year = {2015},
doi = {https://doi.org/10.1103/PhysRevC.91.054317},
author = {Lesher, S. R. and Casarella, C. and Aprahamian, A. and Crider, B. P. and Ikeyama, R. and Marsh, I. R. and  McEllistrem, M. T. and Peters, E. E. and Prados-Estevez, F. M. and Smith, M. K. and Tully, Z. R. and Vanhoy, J. R. and Yates, S. W.},
}

@Article{Bonatsos2024,
AUTHOR = {Bonatsos, Dennis and Martinou, Andriana and Peroulis, Spyridon K. and Petrellis, Dimitrios and Vasileiou, Polytimos and Mertzimekis, Theodoros J. and Minkov, Nikolay},
TITLE = {Robustness of the \textsc{P}roxy-$\textsc{SU}(3)$ Symmetry in Atomic Nuclei and the Role of the Next-Highest-Weight Irreducible Representation},
JOURNAL = {Symmetry},
VOLUME = {16},
YEAR = {2024},
NUMBER = {12},
pages= {1625},
ISSN = {2073-8994},
}

@article{Vargas2013,
  title = {Microscopic study of neutron-rich dysprosium isotopes},
  author = {Vargas, Carlos E. and Velázquez, Víctor and Lerma, Sergio},
  journal = {The European Physical Journal A},
  volume = {49},
  pages = {4},
  year = {2013}
}

@article{Rouoof2024,
  title = {Fingerprints of the triaxial deformation from energies and $\mathit{B}(\mathit{E}2)$ transition probabilities of $\gamma$-bands in transitional and deformed nuclei},
  author = {Rouoof, S. P. and Nazir, Nazira and Jehangir, S. and Bhat, G. H. and Sheikh, J. A. and Rather, N. and Frauendorf, S.},
  journal = {The European Physical Journal A},
  volume = {60},
  pages = {40},
  year = {2024}
}

@article{Jones2025,
  title = {Measurements of octupole collectivity in $^{144}$\textsc{B}a},
  author = {B. Jones},
  journal = {The 29th International Nuclear Physics Conference (INPC 2025)},
  year = {2025}
}

@article{Butler1996,
  title = {Intrinsic reflection asymmetry in atomic nuclei},
  author = {Butler, P. A. and Nazarewicz, W.},
  journal = {Rev. Mod. Phys.},
  volume = {68},
  issue = {2},
  pages = {349--421},
  numpages = {0},
  year = {1996},
  month = {Apr},
  publisher = {American Physical Society},
  doi = {10.1103/RevModPhys.68.349},
  url = {https://link.aps.org/doi/10.1103/RevModPhys.68.349}
}

@article{Butler2016,
doi = {10.1088/0954-3899/43/7/073002},
url = {https://doi.org/10.1088/0954-3899/43/7/073002},
year = {2016},
month = {jun},
publisher = {IOP Publishing},
volume = {43},
number = {7},
pages = {073002},
author = {Butler, P A},
title = {Octupole collectivity in nuclei},
journal = {Journal of Physics G: Nuclear and Particle Physics}
}

@article{Neergard1970,
title = {Low-lying octupole states of the doubly even deformed nuclei with 152$\leqslant{A}\leqslant$190},
journal = {Nuclear Physics A},
volume = {145},
number = {1},
pages = {33-80},
year = {1970},
issn = {0375-9474},
doi = {https://doi.org/10.1016/0375-9474(70)90309-X},
url = {https://www.sciencedirect.com/science/article/pii/037594747090309X},
author = {K. Neergard and P. Vogel}
}

@book{Bohr_Mottelson,
title = {Nuclear Structure},
publisher = {Benjamin, New York},
volume = {II},
number = {},
pages = {},
year = {1975},
issn = {},
doi = {},
url = {},
author = {A. Bohr and B.R. Mottelson}
}

@article{Otsuka2025,
title = {Prevailing triaxial shapes in atomic nuclei and a quantum theory of rotation of composite objects},
journal = {The European Physical Journal A},
volume = {61},
number = {5},
pages = {126},
year = {2025},
issn = {1434-601X},
doi = {10.1140/epja/s10050-025-01553-1},
url = {https://doi.org/10.1140/epja/s10050-025-01553-1},
author = {Otsuka, T. and Tsunoda, Y. and Shimizu, N. and Utsuno, Y. and Abe, T. and Ueno, H.}
}

@article{Kibedi2002,
title = {Reduced electric-octupole transition probabilities, $\mathit{B}(\mathit{E}3;0_1^+ \rightarrow3_1^{-})$-an update},
journal = {Atomic Data and Nuclear Data Tables},
volume = {80},
number = {1},
pages = {35-82},
year = {2002},
issn = {0092-640X},
doi = {https://doi.org/10.1006/adnd.2001.0871},
url = {https://www.sciencedirect.com/science/article/pii/S0092640X0190871X},
author = {Kib\'{e}di, T. and Spear, R. H.}
}

@article{Cottle1990,
  title = {New ``Octupole-driving particle numbers'' from examination of ${3}_{1}^{\mathrm{\ensuremath{-}}}$ state energies},
  author = {Cottle, P. D.},
  journal = {Phys. Rev. C},
  volume = {42},
  issue = {4},
  pages = {1264--1266},
  numpages = {0},
  year = {1990},
  month = {Oct},
  publisher = {American Physical Society},
  doi = {10.1103/PhysRevC.42.1264},
  url = {https://link.aps.org/doi/10.1103/PhysRevC.42.1264}
}

@misc{Nudat,
  title = {NUDAT 3.0},
  author= {},
  year = {2025},
  howpublished = {https://www.nndc.bnl.gov/nudat3/}
}

@article{Chen2021,
  title = {Microscopic origin of reflection-asymmetric nuclear shapes},
  author = {Chen, Mengzhi and Li, Tong and Dobaczewski, Jacek and Nazarewicz, Witold},
  journal = {Phys. Rev. C},
  volume = {103},
  issue = {3},
  pages = {034303},
  numpages = {10},
  year = {2021},
  month = {Mar},
  publisher = {American Physical Society},
  doi = {10.1103/PhysRevC.103.034303},
  url = {https://link.aps.org/doi/10.1103/PhysRevC.103.034303}
}

@article{Martinou2024,
doi = {10.1088/1402-4896/ad562f},
url = {https://doi.org/10.1088/1402-4896/ad562f},
year = {2024},
month = {jun},
publisher = {IOP Publishing},
volume = {99},
number = {7},
pages = {075311},
author = {Martinou, Andriana and Minkov, Nikolay},
title = {Microscopic derivation of the octupole magic numbers from symmetry considerations},
journal = {Physica Scripta}
}

@article{Bizzeti2004,
  title = {Description of nuclear octupole and quadrupole deformation close to the axial symmetry and phase transitions in the octupole mode},
  author = {Bizzeti, P. G. and Bizzeti-Sona, A. M.},
  journal = {Phys. Rev. C},
  volume = {70},
  issue = {6},
  pages = {064319},
  numpages = {18},
  year = {2004},
  month = {Dec},
  publisher = {American Physical Society},
  doi = {10.1103/PhysRevC.70.064319},
  url = {https://link.aps.org/doi/10.1103/PhysRevC.70.064319}
}

@article{Garrett2001,
doi = {10.1088/0954-3899/27/1/201},
url = {https://doi.org/10.1088/0954-3899/27/1/201},
year = {2001},
month = {jan},
publisher = {},
volume = {27},
number = {1},
pages = {R1},
author = {P E Garrett},
title = {Characterization of the $\beta$ vibration and 0$^+_2$ states in
deformed nuclei},
journal = {Journal of Physics G: Nuclear and Particle Physics}
}

@article{Aprahamian2025,
title = {The nature of 0$^{+}$ excitations in deformed nuclei},
journal = {Progress in Particle and Nuclear Physics},
volume = {143},
pages = {104173},
year = {2025},
issn = {0146-6410},
doi = {https://doi.org/10.1016/j.ppnp.2025.104173},
url = {https://www.sciencedirect.com/science/article/pii/S0146641025000201},
author = {Ani Aprahamian and Kevin Lee and Shelly R. Lesher and Roelof Bijker}
}

\end{document}